\documentclass[12pt,a4paper]{article}

\usepackage{amsmath}
\usepackage{amssymb,upgreek}
\usepackage{mathrsfs}
\usepackage{latexsym} 
\usepackage{amsfonts}
\usepackage{framed}
\usepackage[colorlinks]{hyperref}
\usepackage{here}
\usepackage[dvipsnames]{xcolor}
\usepackage{graphicx}
\graphicspath{{./input/}}

\newtheorem{theorem}{Theorem}[section]
\newtheorem{remark}[theorem]{Remark}

\newcommand{\rem}[1]{}
\newcommand{\de}{{\rm d}}
\newcommand{\bu}{{\boldsymbol{u}}}
\newcommand{\bv}{{\boldsymbol{v}}}
\newcommand{\bw}{{\boldsymbol{w}}}
\newcommand{\bkappa}{{\boldsymbol{\kappa}}}
\newcommand{\bn}{{\mathbf{n}}}
\newcommand{\bM}{{\mathbf{M}}}

\newcommand{\bP}{{\boldsymbol{\cal P}}}

\newcommand{\bx}{{\boldsymbol{x}}}
\newcommand{\bnu}{{\boldsymbol{\nu}}}
\newcommand{\bmu}{{\boldsymbol{\mu}}}
\newcommand{\bxi}{{\boldsymbol{\xi}}}

\newcommand{\bgamma}{{\boldsymbol{\gamma}}}
\newcommand{\bgammap}{{\boldsymbol{\gamma}_{\scriptscriptstyle\|}}}
\newcommand{\bOmega}{{\boldsymbol{\Omega}}}
\newcommand{\bomega}{{\boldsymbol{\omega}}}

\newcommand{\MH}{{\text{\tiny MH}}}
\newcommand{\RR}{{\textsf{\tiny RR}}}
\newcommand{\bh}{{\mathbf{h}}}
\newcommand{\bA}{{\boldsymbol{\cal A}}}
\newcommand{\barA}{{\,\bar{\!\boldsymbol{\cal A}\,}\!}}
\newcommand{\barJ}{{\,\bar{\!\boldsymbol{\cal J}\,}\!}}
\newcommand{\bE}{{\boldsymbol{\cal E}}}
\newcommand{\bJ}{{\boldsymbol{\cal J}}}
\newcommand{\bR}{{\mathbf{R}}}
\newcommand{\bB}{{\mathbf{B}}}
\newcommand{\bN}{{\mathbf{N}}}
\newcommand{\bfi}{\bfseries\itshape}
\renewcommand{\contentsname}{}

\newcommand{\comment}[1]{\vspace{5mm}\par
\framebox{\begin{minipage}[c]{.95\textwidth} \tt\bfi #1
\end{minipage}}\vspace{5 mm}\par}

\newcommand{\beq}{\begin{equation}}
\newcommand{\eeq}{\end{equation}}

\begin{document}

\title{Ideal interaction dynamics of  defects\\and background fields in condensed matter
}

\author{
	Fran\c{c}ois Gay-Balmaz$^1$, Tudor S. Ratiu$^{1,2,3}$,  Cesare Tronci$^{1,4}$
	\smallskip\\
	\small
	$^1$\it School of Physical and Mathematical Sciences, Nanyang Technological University, Singapore\\
	\small
	$^2$\it School of Arts and Sciences, Fuyao University of Science and Technology, Fuzhou, China\\
	\small
	$^3$\it  Section de Math\'ematiques, \'Ecole Polytechnique F\'ed\'erale de Lausanne, Lausanne, Switzerland\\
	\small
	$^4$\it School of Mathematics and Physics, University of Surrey, Guildford, United Kingdom
}
\date{}

\maketitle
\begin{abstract}\vspace{-.73cm}
While defect dynamics in condensed matter systems is widely studied and several methods are available to track the defect position over time, the formulation of dynamical models for the coupled evolution of both defects and background field quantities seems to be more challenging. Yet, the overall behavior of a system containing moving defects is strongly influenced by this defect-field interaction. Here, we propose a modeling framework for the conservative coupled evolution of the defect density and the background fields upon focusing on two relevant cases of study: the Gross-Pitaevskii theory of quantum fluids and the Ericksen-Leslie model of nematic liquid crystals. To achieve the coupled evolution, we combine the   Euler-Poincar\'e variational setting of  continuum dynamics with the gauge theory of topological defects. While the latter allows us to overcome certain ambiguities in definition of the defect density,
the Euler-Poincar\'e method provides a unified modeling approach  in the absence of defect  reconnection. Suitable variational regularizations are also proposed to deal with the motion of singular defect lines.
\end{abstract}

\vspace{-1.35cm}

{\small
\contentsname
\tableofcontents
}


\section{Introduction}

The dynamics of topological defects in condensed matter has a crucial influence on key material properties, enabling new functionalities in emerging technologies. In complex fluids, such as liquid crystals, defects are engineered to control self-assembly and affect flow, offering possibilities for tailoring materials with specific properties. Likewise, the appearance of vortex singularities  crucially affects the hydrodynamic properties of quantum fluids, which play a promising role in ultra-precise measurements and new forms of computing. Nevertheless, our ability to predict and control defect–fluid interactions remains limited.

Most existing theories describe defect motion within a quasistatic picture in which defects move through a motionless medium. Valid for highly dissipative regimes, this approximation breaks down in the presence of inertial effects that amplify the defect coupling to the bulk. Also, the recent interest in active liquid crystals comprising self-propelled particles has prompted new approaches to defect-fluid coupling which, however,  combine ad-hoc patches from earlier models. Inertial effects are also prominent in quantum fluid turbulence, where the interaction between vortex singularities and the fluid motion becomes unavoidable.

To uncover universal principles of defect–fluid coupling, here we focus on two complementary systems: (1) quantum vortices in quantum fluids, and (2) disclination defects in nematic liquid crystals. 

In the first case, the Gross-Pitaevskii equation governing Bose-Einstein condensates is known to develop singular vortex structures that are well known in standard quantum theory \cite{Berry,  Birula, Dirac, Barenghi,Berloff,Takabayasi}. Their dynamics is very intricate since it generally involves topological changes occurring during their appearance, disappearance, and reconnection. Even neglecting these effects, the ideal Biot-Savart model \cite{Nazarenko} of quantum vortices generally disregards the evolution of the overall system wavefunction. A very first attempt to bring together vortex location and wavefunction dynamics is due to Takabayasi \cite{Takabayasi}, who extended Dirac's earlier construction \cite{Dirac} to the case of time-dependent wavefunctions by using Madelung's  hydrodynamic formulation \cite{Madelung}. In this treatment, the overall phase gradient is split as $\nabla S (\bx, t)= \nabla S_{reg}(\bx, t) + \bA(\bx)$, 
that is, the sum of a regular part $\nabla S_{reg}$ and a nonintegrable part given by a gauge vector potential $\bA$ \cite{Dirac}. The vortex line is introduced explicitly by writing $\operatorname{curl}\bA = \Gamma {\int}\bR_s \delta(\bx-\bR)\,\de s$, where $\Gamma$ is the topological charge, $\bR( s)$ is the vortex line, and we denote $\bR_s=\partial_s\bR$. Then, resorting to covariant derivatives by the replacement $-i\hbar\nabla\rightsquigarrow -i\hbar\nabla+\hbar\bA$ leads to the nonlinear Schr\"odinger equation 
\beq\label{covarNLS}
i\hbar\frac{\partial\psi}{\partial t}=\frac{\hbar^2}{2m}({-}i\nabla+\bA)^2\psi+(V-\hbar\Phi)\psi+g|\psi|^2\psi, 
\eeq
where we have also accounted for the scalar gauge potential $\Phi$.
As usual, $m$ is the mass, $V(\bx)$ denotes the external potential, and $g$ is a fixed parameter.
In this formulation, however, the vortex position is time-independent and thus it can only be used to model the presence of  a static delta-like magnetic field $\operatorname{curl}\bA$, as in the  Aharonov-Bohm effect \cite{Aharonov, Berry,FoTr}. In this paper, we present a geometric variational construction to prescribe the evolution of Dirac's vector potential $\bA$ in such a way to model the  compound Biot-Savart-Schr\"odinger dynamics of the overall wavefunction $\psi$ and the vortex  line $\bR$. In particular, we will modify the usual Dirac-Frenkel variational principle for Schr\"odinger dynamics by introducing  the vortex velocity as an additional variable, which  will then be related to the Madelung hydrodynamic velocity after taking variations. When completing the Madelung transform so that the quantum fluid has itself a hydrodynamic formulation, the resulting \emph{Madelung vortohydrodynamics} model couples the quantum fluid dynamics to the evolution of the quantum vorticity.

In further developing the proposed modeling framework, we will extend the above construction to consider systems with symmetry breaking and, in particular, we will consider Volovik's variant of the conservative Ericksen-Leslie theory of nematic liquid crystal flows \cite{Volovik}. We will focus on disclination lines, that are defects associated to the orientational order of  the nematic molecules of the  liquid.
In this case,  coupling the defect motion to the other dynamical fields is a much more challenging task, mostly due to the symmetry breaking. The challenge is further complicated by the apparently contrasting definitions of  disclination density appearing in the literature. For example, Kl\'eman's curvature tensor definition \cite{Kleman}, also appearing in Volovik's work \cite{Volovik}, is  very different  from Halperin's definition \cite{Halperin}, which  is identified by a singular vector field. Also, in some  cases, the disclination density is identified with $\operatorname{curl}\bv_\MH=\epsilon_{abc}n^a\nabla n^b\times \nabla n^c/2$, where $\bv_\MH$ is the Mermin-Ho velocity from the superfluid theory of Helium-3 \cite{MerminHo} and $\bn(\bx,t)$ is the order parameter field such that $|\bn(\bx,t)|=1$. With such differences appearing already in the fundamental works, a systematic study of defect dynamics necessitates establishing relations among the different definitions. Luckily, as shown in this work, an answer in this sense  is provided by Eringen's micropolar formulation of liquid crystal dynamics \cite{Eringen}. As unfolded by the authors in \cite{GBRaTr2012,GBRaTr2013}, a particular role is played by the geometry underlying the wryness tensor $\bgamma$ satisfying 
\beq
\partial_j \bn=\bn\times \bgamma_j.
\eeq 
As we shall see, this quantity leads to the establishment of an unambiguous definition of the disclination density. However, in coupling disclinations to the background fields,  further challenges are due to the fact that the disclination density does not appear explicitly in the  Ericksen-Leslie model, which  indeed is known to fail in capturing disclination dynamics \cite{Cermelli}. To overcome these challenges, we will take suitable modeling steps thereby realizing the coupling of the disclination dynamics to both the nematic texture and the bulk flow. In the case of singular disclination lines, we will take advantage of the variational structure underlying the continuum theory to provide variational regularization models tracking the interaction dynamics of the defect with the surrounding medium.

As stated above, the proposed models focus on the ideal reversible dynamics of conservative condensed matter systems. In the case of quantum fluids, this means that the model does not capture the spontaneous appearance of vortex structures nor their reconnection after collisions. Similar limitations apply in the case of liquid crystal flows. More importantly, in the latter case, we  focus on rotation inertial effects which are usually discarded in real-world scenarios. Active nematics and liquid crystal turbulence do instead involve relevant inertial effects \cite{Koch}. Despite these limitations, we believe that a complete description advancing the coupled dynamics of disclinations and background fields represents a step forward, not only in our understanding of the mathematical problems underlying condensed matter systems \cite{Zarnescu}, but also in the possibility to understand physical coupled phenomena that presently lack a theoretical formulation.

\paragraph{Plan of the paper.} Section \S\ref{sec:quantvort} extends the Dirac-Takabayasi construction for constant vortex locations to the time-dependent case by applying minimal coupling to the Dirac-Frenkel variational principle underlying the nonlinear Schr\"odinger equation. As a result, a new model is presented for the coupled dynamics of both vorticity and quantum wavefunction. This construction is specialized to the case of singular vortex filaments in \S\ref{sec:vortlines}, where a variational regularization allows the establishment of a coupled Biot-Savart-Schr\"odinger system for the coupled dynamics of vortex lines and quantum wavefunctions. Subsequently, the Madelung transform is applied to the wavefunction in \S\ref{sec:MVHD}, thereby leading to a coupled hydrodynamic system of vorticity and Madelung flow. A direct analogy of the resulting system with Hall magnetohydrodynamics is also briefly discussed. Later on, \S\ref{sec:MCHDHamSt} presents the explicit Hamiltonian structure of the coupled systems appeared so far. Section \S\ref{sec:DisclText} moves on to apply the general construction to the case of nematic textures in the absence of hydrodynamic flow. After a brief introduction to the dynamics in the absence of defects in \S\ref{sec:DisclTextSyst}, an apparent ambiguity involving different disclination density definitions is resolved in \S\ref{sec:discldens}. In \S\ref{sec:coupleddisclnem}, this result allows to apply the general minimal coupling construction to present a new model for the coupled dynamics of disclination density and the director field, along with its angular momentum. We show that the parallel component of the angular momentum along the director field is not allowed to vanish anywhere in the presence of disclinations. While \S\ref{sec:HamDisclNem} presents both the Lagrangian and the Hamiltonian formulations, \S\ref{sec:BS-RR}  presents a variational regularized model for the coupled dynamics of  a singular disclination line interacting with the surrounding  nematic texture.  The discussion proceeds in \S\ref{sec:liqxtals} by considering the case of flowing nematic liquid crystals. The coupled system of disclination density dynamics and liquid crystal motion is presented in \S\ref{sec:LiqxtalEqns}, along with a regularized model to track singular disclination lines, whereas the Lagrangian and Hamiltonian formulations are discussed in \S\ref{sec:LiqxtalEqnsHam}.

\section{Ideal vorticity dynamics in quantum fluids\label{sec:quantvort}}

In this section, we will extend the nonlinear Schr\"odinger equation \eqref{covarNLS}  and its Madelung hydrodynamic variant in such a way that Dirac's vector potential $\bA$ has nontrivial dynamics. Eventually, we will achieve an evolution model for the coupled dynamics of  the quantum vorticity 
\[
\bomega=\operatorname{curl}\bA
\] 
and the state of the quantum fluid.

\subsection{Dirac-Frenkel variational principle and vortex dynamics}

 As anticipated above, we modify the standard Dirac-Frenkel variational principle to include the vortex motion as encoded in $\bA(\bx,t)$. To this purpose, we propose the following variational principle:
\begin{multline}\label{VarPr1}
\delta{\int_{t_1}^{t_2}}{\int}\bigg[{\operatorname{Re}}\bigg(i\hbar\psi^*\frac{\partial\psi}{\partial t}\bigg)-\frac{1}{2m}\psi^*(i\hbar\nabla - \hbar\bA)^2\psi-\bigg(V-\hbar\Phi+\frac{g}2|\psi|^2\bigg)|\psi|^2
\\
+\sigma (\bA\cdot \bv-\Phi)-\frac{1}2\bomega(-\Delta^{-1})\bomega
\bigg]\de^3x\de t=0,
\end{multline}
where  $\sigma(\bx,t)$ is the  gauge charge density, $\Phi(\bx, t)$ is the scalar potential, and Planck's constant $\hbar$  plays the role of a coupling constant associated to the gauge fields \cite{Dirac,FoHoTr19}. The presence of the minimal-coupling gauge term $\sigma (\bA\cdot \bv-\Phi)$  allows for time-dependent vortex motion in such a way that, as we shall see, the velocity  field $\bv(\bx,t)$ identifies the vortex velocity. Also,
following Kleinert's work \cite{Kleinert,Kleinert1},   the last term in \eqref{VarPr1} 
represents the Biot-Savart vortex interaction energy. The necessity of such a term will become manifest later on.

The Schr\"odinger equation associated to  \eqref{VarPr1} results from  the variations $\delta\psi$, that is,
\[
i\hbar\frac{\partial \psi}{\partial t}=-\frac{\hbar^2}{2m}(\nabla +i \bA)^2\psi+(V+g|\psi|^2-\hbar\Phi)\psi,
\]
while  arbitrary variations $\delta \Phi$ and $\delta\bA$ yield 
\beq\label{sigma-vIds}
\sigma=\hbar|\psi|^2
\,,\qquad\qquad
\bv=\frac{\hbar^2}{m\sigma}\big(\operatorname{Im}(\psi^*\nabla\psi)+|\psi|^2\bA\big)+\frac1\sigma\bJ
,
\eeq
where we have introduced the notation
\beq\label{vortcorr}
\bJ=
-\operatorname{curl}\Delta^{-1}\bomega.
\eeq
The equation for $\bA(\bx, t)$, instead, is given by the variations $\delta\bv$ and $\delta\sigma$ which need extra care. As anticipated, in this construction $\bv(\bx,t)$ plays the role of a hydrodynamic vortex velocity and is defined in terms of the diffeomorphic Lagrangian path $\boldsymbol\eta(\bx_0,t)$ as $\dot{\boldsymbol\eta}(\bx_0,t)=\bv(\bx,t)|_{\bx=\boldsymbol\eta(\bx_0,t)}$.  Analogously, the charge density $\sigma(\bx,t)$ is prescribed to be transported by the flow as $\sigma(\bx,t)={\int}\sigma_0(\bx_0)\delta(\bx-\boldsymbol\eta(\bx_0,t))\de^3x_0$, so that its evolution is prescribed as
\beq\label{GCT}
\partial_t\sigma+ \operatorname{div}(\sigma \bv)=0.
\eeq
Then, as standard in the Euler-Poincar\'e theory of continuum media \cite{HoMaRa98}, the variations $\delta\bv$ and $\delta\sigma$ are evaluated in terms of the arbitrary variation $\delta\boldsymbol\eta$ such that $\delta\boldsymbol\eta(\bx_0,t_1)=\delta\boldsymbol\eta(\bx_0,t_2)=0$.
Upon introducing the arbitrary quantity $\bw(\bx,t)=\delta{\boldsymbol\eta}(\bx_0,t)|_{\bx_0=\boldsymbol\eta^{-1}(\bx,t)}$, one has
\beq\label{vars1}
\delta\bv=\partial_t \bw+\bv\cdot\nabla\bw-\bw\cdot\nabla\bv
\,,\qquad\qquad
\delta\sigma=-\operatorname{div}(\sigma\bw),
\eeq
thereby yielding $\partial_t\bA-\bv\times \operatorname{curl}\bA=-\nabla \Phi$.

Therefore, we obtain the coupled system
\begin{align}\label{NewSchrod}
&i\hbar\frac{\partial \psi}{\partial t}=-\frac{\hbar^2}{2m}(\nabla +i \bA)^2\psi+(V+g|\psi|^2-\hbar\Phi)\psi,
\\\label{ohmslaw}
&\frac{\partial \bA}{\partial t}+\nabla \Phi=\bv\times \operatorname{curl}\bA
\end{align}
with
\beq\label{vortvel}
\bv
=\frac{\hbar}{m|\psi|^2}\big(\operatorname{Im}(\psi^*\nabla\psi)+|\psi|^2\bA\big)+\frac1{\hbar|\psi|^2}\bJ
.
\eeq
Notice that, since $\hbar^{-1}\sigma/|\psi|^2=1$, the last term in the above expression of the vortex velocity $\bv$ does not contribute to the divergence in gauge charge transport \eqref{GCT}, which then recovers the continuity equation for the probability density $|\psi|^2$.

We recognize that the equations \eqref{NewSchrod}-\eqref{vortvel}  represent a closed system that involves solving directly for the gauge potentials. For example, one can  use the Coulomb gauge $\operatorname{div}\bA=0$ as in Takabayasi's work \cite{Takabayasi}, so that $\bA=-\operatorname{curl}\Delta^{-1}\bomega$ and $\Delta \Phi=\operatorname{div}(\bv\times \bomega)$,  and this gauge choice is the focus of the next section.

\subsection{Regularized Biot-Savart-Schr\"odinger model of  vortex dynamics\label{sec:vortlines}}

In this section, we shall consider the dynamics of singular vortex filaments. In this context,  Takabayasi's choice of the Coulomb gauge allows for an explicit model 
in which the Biot-Savart vortex dynamics is coupled to the nonlinear Schr\"odinger equation \eqref{NewSchrod} for the quantum fluid evolution. 
In Takabayasi's original paper \cite{Takabayasi},  the vortex location $\bR(s,t)$ is inserted via 
\beq\label{vortex}
\bomega(\bx, t)=\Gamma {\int}\bR_s(s,t) \delta(\bx-\bR(s,t))\,\de s,
\eeq
where $\Gamma=2n\pi\hbar/m$ and $n$ is an integer. Here, the Coulomb gauge allows to apply the Biot-Savart law to write $\bA=\operatorname{curl}^{-1}\bomega$, where the operator $\operatorname{curl}^{-1}=-\operatorname{curl}\Delta^{-1}$ selects the unique divergence-free vector potential that vanishes at infinity. In this case, one verifies that \eqref{vortex} implies
\begin{equation}\label{A_t}
-\partial_t\bA-\nabla\Phi=\bE,
\end{equation}
with
\beq\label{elfield}
\Phi= -\operatorname{div}\Delta^{-1}\bE
\,,\qquad\text{ and }\qquad
\bE= \Gamma {\int}\bR_s\times {\bR}_t\,\delta(\bx-\bR)\,\de s.
\eeq
At this point, as usual in vortex filament dynamics, these singular expressions lead to  the blowup of our model equations. For example, as  in the case of  Euler incompressible fluids, replacing \eqref{vortex} in the conserved energy functional
\beq\label{TotEn1}
\mathsf{H}(\psi, \bA)={\int}\bigg(\frac{1}{2m}\psi^*(i\hbar\nabla - \hbar\bA)^2\psi+\bigg(V+\frac{g}2|\psi|^2\bigg)|\psi|^2
+\frac{1}2\bomega\cdot (-\Delta^{-1})\bomega
\bigg)\de^3x
\eeq
leads to a divergent integral. 

To make further progress and arrive at a well-defined final set of equations, a regularization becomes necessary and indeed different regularization strategies are available. 
Here, we replace the gauge potentials $(\Phi,\bA)$ by their convolution $(\bar\Phi,\barA)=(K*\Phi,K*\bA)$ with a translation-invariant mollifier $K(\bx,\bx')=K(\bx-\bx')$ such that $K(-\bx)=K(\bx)$ and ${\int} K(\bx-\bx')\,\de^3 x=1$; e.g., a Gaussian kernel. Similar regularizations have also been proposed in computational chemistry \cite{RaTr20} to deal with molecular topological phases \cite{Mead}. In the present case, the regularization step amounts to suitably replacing $(\Phi,\bA)\rightsquigarrow(\bar\Phi,\barA)$ in the original variational principle \eqref{VarPr1}, which then becomes
\begin{multline}\label{VarPr1reg}
\delta{\int_{t_1}^{t_2}}{\int}\bigg({\operatorname{Re}}\big(\psi^*(i\hbar\partial_t+\hbar\bar{\Phi})\psi\big)
-\frac1{2m}\psi^*|i\hbar\nabla -\hbar \barA|^2\psi-\Big(V+\frac{g}2|\psi|^2\Big)|\psi|^2
\\
+\sigma (\bA\cdot \bv-\Phi)
-\frac12\bar\bomega\cdot(-\Delta^{-1})
                    \bar\bomega
\bigg)\de^3x\de t=0.
\end{multline}
We observe that, while the minimal coupling term in the second line was left unchanged, the replacement $\bA\rightsquigarrow\barA$ affected the Biot-Savart energy. In addition, we notice that this regularization ensures that \eqref{VarPr1reg} is invariant under the modified gauge transformation
\beq\label{reggauge}
\bA\mapsto \bA+\nabla \varphi
,\qquad\qquad
\Phi\mapsto \Phi-\partial_t \varphi
,\qquad\qquad
\psi\mapsto 
\psi\exp(-i\bar\varphi),
\eeq
for a fixed kernel $K$. Here, the potentials transform as usual and the regularization affects only the action on the wavefunction; we will refer to \eqref{reggauge} as \emph{regularized gauge transformation}.
The associated gauge symmetry is crucial to ensure that the constraint 
\beq\label{QNreg}
\sigma=\hbar\overline{\,|\psi|^2},
\eeq
resulting from the variations $\delta\Phi$, is preserved by the dynamics via Noether's theorem. Notice that the notation is such that the overbar always denotes convolution with the same translation-invariant mollifier $K$. The variations $\delta\bA$ in \eqref{VarPr1reg} yield
\beq\label{vortvelreg}
\bv=\frac{\hbar^2}{m\sigma}\overline{\bigg({\operatorname{Im}}(\psi^*\nabla\psi)+|\psi|^2\barA+\frac{m}{\hbar^2}\barJ\bigg)} 
.
\eeq
Then, taking the variations $\delta\bv$ and $\delta\sigma$ recovers \eqref{ohmslaw} while the variations $\delta\psi$ yield the regularized Schr\"odinger equation
\beq\label{NewSchrodreg}
i\hbar\frac{\partial\psi}{\partial t}
=-\frac{\hbar^2}{2m}
 (\nabla+i\overline{\bA})^2\psi
+\left(V+g|\psi|^2-\hbar\overline\Phi\right)\psi.
\eeq
Notice that, since ${\operatorname{div}}\barJ=0$, we have  $\hbar\partial_{t\!}{\overline{\,|\psi|^2}}=-\hbar^{2} m^{-1}{\operatorname{div}}\overline{({\operatorname{Im}}(\psi^*\nabla\psi)+|\psi|^2\barA)}
=-
\operatorname{div}(\sigma\bv)=\partial_t\sigma$, so that the constraint \eqref{QNreg} is preserved as anticipated.

At this point, one introduces the vortex filament \eqref{vortex} and  resorts to the Coulomb gauge so that  ${\operatorname{div}}\barA=\overline{{\operatorname{div}}\bA}=0$ and $\barA=\operatorname{curl}^{-1}\bar\bomega=\barJ$. Then, inserting  \eqref{A_t} in ${\partial_t \bA}+\nabla \Phi={\bv}\times \operatorname{curl}\bA$ gives the ideal Ohm's law  $\bE+\bv\times\bomega=0$, thereby leading to a  regularized version of the usual Magnus force relation \cite{Barenghi,Sonin}, that is
\beq\label{Magnus}
\bR_s\times({\bR}_t-{\bv}(\bR))=0,
\qquad\text{where}\qquad
\bv(\bR)=\left.\frac\hbar{m\overline{\,|\psi|^{2}}}\overline{\bigg({\operatorname{Im}}(\psi^*\nabla\psi)+|\psi|^2\barJ+\frac{m}{\hbar^{2}}
\barJ\bigg)}\right|_{\bx=\bR}
\eeq
 is   the regularized vortex velocity in \eqref{vortvelreg}, after using \eqref{QNreg}.  Finally, recalling \eqref{elfield} we rewrite the regularized Schr\"odinger equation \eqref{NewSchrodreg} in the form 
\beq\label{NewSchrodreg2}
i\hbar\frac{\partial \psi}{\partial t}=-\frac{\hbar^2}{2m}(\nabla +i \barJ)^2\psi+(V+g|\psi|^2+\hbar\operatorname{div}\Delta^{-1}\bar\bE)\psi,
\eeq
Thus, equations \eqref{Magnus} and \eqref{NewSchrodreg2}  comprise a Hamiltonian \emph{Biot-Savart-Schr\"odinger} system which extends a similar model previously appeared in \cite{FoTr} to account for a relative vortex velocity.

As an alternative to the Coulomb gauge, when approaching the entire PDE system \eqref{NewSchrod}-\eqref{vortvel} for a smooth vortex distribution, it may also be  convenient to choose Weyl's temporal gauge $\Phi=0$ thereby eliminating one variable. For example, this choice is convenient in the formulation of the underlying Hamiltonian structure, as shown  later on.

In the next section, we show how Madelung's hydrodynamic formulation can be used to eliminate the gauge potentials in such a way to express the dynamics in terms of the vortex density $\bomega$.


\begin{remark}[Conservation of vortex helicity]\rm Notice that, while  this treatment can be easily extended to multiple vortex filaments by writing $\bomega=\sum_j\Gamma_j \int\bR_{j s} \delta(\bx-\bR_j)\,\de s$, the vortex reconnection normally associated to their collisions \cite{Berry1,Ricca,Zuccher} is not captured by this model. Indeed, the present setting does not capture topological changes and this is also seen in the absence of singularities due to the conservation of the
the vortex helicity 
\beq\label{vorthel}
{\cal H}_{\rm v}={\int}\bomega\cdot\bA\,\de^3x.
\eeq 
The conservation law for this topological quantity is obtained by taking the curl of \eqref{ohmslaw}. 
Therefore, the present ideal model fails to track topological changes and the vortex dynamics is  treated  in a pure hydrodynamic setting. Unlike Euler hydrodynamics, however, the vortex velocity is allowed here to be a compressible vector field. At a computational level, vortex reconnection may be included  by adopting the numerical patches currently used in superfluid simulations, in such a way that the present model is implemented to run before and after reconnection  events.
\end{remark}

\subsection{Madelung vortohydrodynamics and its variational structure\label{sec:MVHD}}

In this section, we show how the Madelung hydrodynamic formulation can be used to simplify the mathematical treatment of the PDE system \eqref{NewSchrod}-\eqref{vortvel} in the case of a smooth vortex distribution. In addition, we will see how the Biot-Savart energy term in the second line of \eqref{VarPr1} is absolutely necessary in realizing the coupled model. As a result, we provide a new model for the interaction dynamics of quantum vorticity and the Madelung hydrodynamics of the quantum fluid. As we will see, the name \emph{Madelung vortohydrodynamics} is inspired by the appearance of close analogies with  magnetohydrodynamics models in the physics of magnetized plasmas \cite{Narita}.

We start by writing the Madelung formulation \cite{Madelung} of the nonlinear Schr\"odinger equation \eqref{NewSchrod}. 
Upon writing $\psi=\sqrt{\rho}e^{iS\hbar}$ and denoting $m\bu=\nabla S+\hbar\bA$,  Takabayasi's equations \cite{Takabayasi} 
read
\begin{align}
&m(\partial_t\bu+\bu\cdot\nabla)\bu=\hbar(\partial_t\bA+\nabla\Phi-\bu\times\bomega)-\nabla (V+V_Q+g\rho),
\\
&\partial_t\rho+\operatorname{div}(\rho\bu)=0,
\end{align}
where 
\[
V_Q=-\frac{\hbar^2}{2m}\frac{\Delta\sqrt{\rho}}{\sqrt{\rho}}
\]
is Bohm's quantum potential and we have restored the scalar gauge potential $\Phi$ for consistency.

Considering our system \eqref{NewSchrod}-\eqref{vortvel}, we observe that \eqref{vortvel} reads
\[
\bv=\bu+\frac1{\hbar\rho}\bJ.
\]
Then, using \eqref{ohmslaw}, one rewrites the system \eqref{NewSchrod}-\eqref{vortvel} in hydrodynamic form as
\begin{align}\label{QFmom}
&m\rho(\partial_t\bu+\bu\cdot\nabla\bu)=\bJ\times\bomega-\rho\nabla (V_Q+V+g\rho),
\\\label{vortdyn}
&\partial_t \bomega=\operatorname{curl}\Big(\bu\times\bomega+\frac{1}{\hbar\rho}\bJ\times \bomega\Big),
\\\label{dens}
&\partial_t\rho+\operatorname{div}(\rho\bu)=0.
\end{align}
We observe that, in this formulation, the gauge potentials are entirely absent: the vorticity is now treated as a dynamical variable in a manifestly gauge invariant fashion. More importantly, we observe that the Biot-Savart vortex energy in the second line of \eqref{VarPr1} is uniquely responsible for the feedback of the vortex dynamics on the quantum fluid flow, which otherwise decouples completely. This explains the necessity of retaining this energy component in the treatment. We also notice that, in the presence of singular vortex filaments, the delta vorticity profile appears explicitly in the hydrodynamic momentum equation \cite{Sonin}, thereby posing formidable challenges in the analysis of the resulting PDE system. As we saw before, the conserved  energy functional \eqref{TotEn1}, now written in terms of $\rho=|\psi|^2$ and $\bu=m^{-1}(\nabla S+\hbar\bA)$ as
\beq\label{totenMadVHD}
\mathsf{H}(\bu,\rho, \bA)={\int}\bigg(
\frac{m}{2}\rho|\bu|^2+\frac{\hbar^2}{8m}\frac{|\nabla\rho|^2}\rho
+V\rho +\frac{g}2\rho^2+ \frac{1}2\bomega(-\Delta^{-1})\bomega\bigg)\de^3x,
\eeq
diverges upon replacing \eqref{vortex}.



For later purpose, here we illustrate the Euler-Poincar\'e variational structure of equations \eqref{QFmom}-\eqref{dens}, which is given by the following variational principle
\begin{multline}\label{MadVHD-vp}
\delta{\int_{t_1}^{t_2}}{\int}\bigg(
\frac{m}{2}\rho|\bu|^2-\frac{\hbar^2}{8m}\frac{|\nabla\rho|^2}\rho
+\left(\sigma\bv-\hbar\rho\bu\right)\cdot\bA -V\rho -\frac12 g\rho^2-\Phi (\sigma-\hbar\rho)
\\
-\frac{1}2(\operatorname{curl}\bA)\cdot (-\Delta^{-1})(\operatorname{curl}\bA)\bigg)\de^3x=0,
\end{multline}
along with the variations \eqref{vars1} and
\beq\label{vars2}
\delta\bu=\partial_t \bkappa+\bu\cdot\nabla\bkappa-\bkappa\cdot\nabla\bu
\,,\qquad\qquad
\delta\rho=-\operatorname{div}(\rho\bkappa).
\eeq
Here, $\bkappa(\bx,t)$ is an arbitrary vector field such that $\delta{\boldsymbol\zeta}(\bx_0,t)=\bkappa(\bx,t)|_{\bx=\boldsymbol\zeta(\bx_0,t)}$, where $\boldsymbol\zeta(\bx_0,t)$ is the Lagrangian path associated to the Madelung  fluid flow  and identifying the so-called Bohmian trajectories \cite{Bohm,FoHoTr19,Wyatt}. 
Notice that these trajectories differ from the Lagrangian paths $\boldsymbol\eta(\bx_0,t)$ associated with the vortex velocity $\bv$. Also, as before,  the variations $\delta \Phi$ impose the relation $\sigma= \hbar \rho$ between the densities associated with each of the flows. The action principle \eqref{MadVHD-vp} represents the Madelung hydrodynamic variant of the  Dirac-Frenkel action principle \eqref{VarPr1} associated to the Scr\"odinger wavefunction picture.

Without external forces and in the absence of the quantum potential, a similar variational principle has been proposed in the physics of magnetized plasmas, specifically for the Hall magnetohydrodynamics model \cite{Holm,Ilgisonis}. Indeed, for a smooth vorticity distribution, the system \eqref{QFmom}-\eqref{dens} is reminiscent of the pressureless equations of Hall magnetohydrodynamics, which are obtained from \eqref{QFmom}-\eqref{dens} upon setting $\hbar=1$, $g=0$, and formally replacing $-\Delta^{-1}\rightsquigarrow\operatorname{Id}$, where $\operatorname{Id}$ denotes the identity operator. In that context, the vorticity $\bomega$ is replaced by a magnetic field and the vortex current $\boldsymbol{\cal J}$ becomes a charge current generating the Hall term $\boldsymbol{\cal J}\times\bomega/\rho$, while the vortex velocity $\bv$ is the electron velocity \cite{Ilgisonis}. While on this occasion we will not expand on this interesting analogy, the formal relation to  Hall magnetohydrodynamics inspires  the name ``Madelung vortohydrodynamics'' for the model  \eqref{QFmom}-\eqref{dens}. The term \emph{vortodynamics} was introduced in the context of quantum vortices in \cite{Fischer}.

Before proceeding to the next section, we observe that the regularized Biot-Savart-Schr\"o\-din\-ger model \eqref{Magnus}-\eqref{NewSchrodreg2} may also be expressed in hydrodynamic form. Indeed, upon denoting $m\bu=\nabla S+\hbar\barA$, the Schr\"odinger equation \eqref{NewSchrodreg} leads to $m(\partial_t\bu+\bu\cdot\nabla)\bu=\hbar(\partial_t\barA+\nabla\bar\Phi-\bu\times\bar\omega-\nabla (V+V_Q+g\rho)$, while  $\bv=\overline{(\rho\bu+{\hbar}^{-1}\barJ)}/\bar\rho$ is the regularized vortex velocity and \eqref{ohmslaw} leads to $\partial_t \barA+\nabla \bar\Phi=\overline{\bv\times \bomega}$.
Then, the Madelung form of the regularized Biot-Savart-Schr\"o\-din\-ger model \eqref{Magnus}-\eqref{NewSchrodreg2} reads
\begin{align*}
&{\partial_t\rho}+\operatorname{div}(\rho\bu)=0,
\\
&m(\partial_t\bu+\bu\cdot\nabla\bu)=\hbar(\overline{\bv\times \bomega}-\bu\times\bar\bomega)-\nabla (V_Q+V+g\rho),
\\
&\bR_s\times({\bR}_t-{\bv}(\bR))=0,
\qquad\text{where}\qquad
\bv(\bR)=\left.\frac1{\bar\rho}\overline{\bigg(\rho\bu+\frac{1}{\hbar}
\barJ\bigg)}\right|_{\bx=\bR}.
\end{align*}
Notice that 
$ 
\overline{\bv\times\boldsymbol\omega}(\bx)
-\bu(\bx)\times\overline{\boldsymbol\omega}(\bx)
=
\int K(\bx-\boldsymbol{y})(\bv(\boldsymbol{y})-\bu(\bx))
\times\boldsymbol\omega(\boldsymbol{y})\,\de^3y$, so that using \eqref{vortex} and the Magnus force equation gives
\[
\overline{\bv\times\boldsymbol\omega}(x)
-\bu(x)\times\overline{\boldsymbol\omega}(x)
=
\Gamma\int K(x-\mathbf{R}(s))
\big(\mathbf{R}_t(s)-\bu(x)\big)
\times\mathbf{R}_s(s)\,\de s,
\]
which is the explicit expression for the regularized feedback force exerted by the vortex on the Madelung flow.
Given the underlying variational structure, this type of Biot-Savart-Madelung regularization may serve as a platform for new structure-preserving simulation codes tracking the interaction of quantum vortices and the superfluid bulk, for example blending recent particle schemes in Madelung hydrodynamics \cite{FoHoTr19,Tr2026}. We leave this direction for future work.


\subsection{Hamiltonian structure\label{sec:MCHDHamSt}}

The Dirac-Frenkel variational principle \eqref{VarPr1} underlying the system \eqref{NewSchrod}-\eqref{vortvel} leads naturally to a Hamiltonian structure that is given, as usual, by a Poisson bracket and a Hamiltonian functional identifying the total energy. In the case of equations \eqref{NewSchrod}-\eqref{vortvel}, this Hamiltonian functional $\mathsf{H}(\psi,\bA)$ is given in first instance  by \eqref{TotEn1}, while the identification of the Poisson bracket structure is the subject of the present section.

As a first step in the derivation of the overall Hamiltonian  structure, we write the variational principle \eqref{VarPr1} for a general Hamiltonian functional $\mathsf{H}( \psi, \bA)$ as follows:
\begin{equation}\label{DF_H}
\delta{{\int_{t_1}^{t_2}}}\bigg[{\int}\bigg({\operatorname{Re}}\bigg(i\hbar\psi^*\frac{\partial\psi}{\partial t}\bigg)+ \sigma\bA\cdot \bv\bigg)\de^3x - \mathsf{H}( \psi, \bA) \bigg]\de t=0,
\end{equation}
for arbitrary variations $\delta \psi$, $\delta \bA$ and Euler-Poincar\'e variations of $(\bv, \sigma)$, where we have chosen the Weyl gauge $\Phi=0$.
The variations $\delta\psi$ and $\delta\bA$ in \eqref{DF_H} yield, respectively,
\[
i\hbar\frac{\partial \psi}{\partial t} = \frac{1}{2} \frac{\delta \mathsf{H}}{\delta \psi},\qquad \frac{\delta \mathsf{H}}{\delta \bA}= \sigma\bv,
\]
while, as we shall see, the condition $\sigma=\hbar|\psi|^2$  is preserved by the dynamics, provided that it holds initially.
In addition, the variations \eqref{vars1} lead to the  equations
\[
\frac{\partial \bP}{\partial t}
+{\operatorname{div}}\bigg(\frac{1}{\sigma}\frac{\delta\mathsf{H}}{\delta \bA}\bP\bigg)
+\nabla\bigg(\frac{1}{\sigma}\frac{\delta\mathsf{H}}{\delta \bA}\bigg)\cdot\bP
=
 \sigma\nabla\bigg(\frac1{\sigma}\frac{\delta\mathsf{H}}{\delta \bA}\cdot{\bP} \bigg)
,\qquad\frac{\partial \sigma}{\partial t}
 + \operatorname{div}\frac{\delta\mathsf{H}}{\delta \bA}=0,
\]
where we have introduced the  the  momentum variable
\beq\label{vortmom}
\bP=\sigma \bA.
\eeq
Then, we allow for an explicit $\sigma$-dependence in the energy functional by defining
\beq
\mathscr{H}(\psi,\bP,\sigma)=\mathsf{H}\bigg(\psi, \frac{\bP}{\sigma}\bigg),
\label{change_variables}
\eeq
so that 
\begin{multline*}
\mathscr{H}(\psi,\bP,\sigma)={\int}\bigg(\frac{1}{2m}\psi^*\bigg(i\hbar\nabla - \frac{\hbar}{\sigma}\bP\bigg)^2\psi+\bigg(V+\frac{g}2|\psi|^2\bigg)|\psi|^2
\\
+\frac{1}2\left(\operatorname{curl}\frac\bP\sigma\right)\cdot(-\Delta^{-1})\left(\operatorname{curl}\frac\bP\sigma\right)
\bigg)\de^3x
\end{multline*}
and
the chain rule ensures
\beq\label{chainrule}
\frac{\delta \mathscr{H}}{\delta \psi}= \frac{\delta \mathsf{H}}{\delta \psi},\qquad\quad \frac{\delta \mathscr{H}}{\delta \bP}= \frac{1}{\sigma}\frac{\delta \mathsf{H}}{\delta \bA}, \qquad\quad \frac{\delta \mathscr{H}}{\delta \sigma}= -\frac{\delta \mathsf{H}}{\delta \bA}\cdot \frac{\bP}{\sigma^2}.
\eeq
The equations resulting from \eqref{DF_H} can thus be rewritten as
\begin{equation}\label{Hamiltonian_system}
i\hbar\frac{\partial \psi}{\partial t} = \frac{1}{2} \frac{\delta \mathscr{H}}{\delta \psi},\qquad 
\frac{\partial \bP}{\partial t}
+{\operatorname{div}}\bigg(\frac{\delta \mathscr{H}}{\delta \bP}\bP\bigg)
+\bigg(\nabla\frac{\delta \mathscr{H}}{\delta \bP}\bigg)\cdot\bP
=
-\sigma\nabla \frac{\delta \mathscr{H}}{\delta \sigma}
,
\qquad\frac{\partial \sigma}{\partial t} + {\operatorname{div}}\bigg(\sigma\frac{\delta \mathscr{H}}{\delta \bP}\bigg)=0.
\end{equation}
At this point, the Hamiltonian structure of the system becomes apparent. In particular, equations \eqref{Hamiltonian_system}, and hence the system \eqref{NewSchrod}-\eqref{vortvel}, are Hamiltonian with respect to the direct-sum Poisson bracket
\begin{multline*}
\{\mathscr{F},\mathscr{H}\}(\psi,\bP,\sigma)=\frac1{2\hbar}\operatorname{Im}{\int}\bigg(\frac{\delta \mathscr{F}}{\delta \psi}\bigg)^{\!*}\frac{\delta \mathscr{H}}{\delta \psi}\,\de^3x
\\
+
{\int}\bP\cdot\bigg(\frac{\delta \mathscr{F}}{\delta \bP}\cdot\nabla\frac{\delta \mathscr{H}}{\delta \bP}-\frac{\delta \mathscr{H}}{\delta \bP}\cdot\nabla\frac{\delta \mathscr{F}}{\delta \bP}\bigg)\de^3 x
-
{\int}\sigma\bigg(\frac{\delta \mathscr{F}}{\delta \bP}\cdot\nabla\frac{\delta \mathscr{H}}{\delta \sigma}-\frac{\delta \mathscr{H}}{\delta \bP}\cdot\nabla\frac{\delta \mathscr{F}}{\delta \sigma}\bigg)\de^3 x.
\end{multline*}
This is a direct-sum Poisson bracket structure given by the sum of the canonical Schr\"odinger
bracket  in $\psi$ (first line, yielding  the first  in \eqref{Hamiltonian_system}) and the hydrodynamic Lie-Poisson bracket  in $(\bP,\sigma)$ (second line, yielding the last two in \eqref{Hamiltonian_system}) underlying compressible barotropic hydrodynamics. 
Lie-Poisson brackets for quantum point vortex dynamics have also appeared recently in \cite{Ohsawa}.

The Hamiltonian structure given above transfers directly to the  Madelung vortohydrodynamics system \eqref{QFmom}-\eqref{dens}, although a convenient change of variables is again  needed. In particular, we introduce the hydrodynamic momentum as it arises from the $\delta\bu$-variations in \eqref{MadVHD-vp}:
\[
\bM=m\rho\bu-\hbar\rho\bA=m\rho\bu-\frac{\hbar\rho}\sigma\bP.
\]
Then, the total energy \eqref{totenMadVHD} leads to the Hamiltonian functional in the form
\begin{multline*}
\mathscr{H}(\bM,\rho,\bP,\sigma)=
{\int}\bigg(
\frac{1}{2m\rho}\bigg|\bM+\frac{\hbar\rho}\sigma\bP\bigg|^2+\frac{\hbar^2}{8m}\frac{|\nabla\rho|^2}\rho+\bigg(V+\frac{g}2|\psi|^2\bigg)|\psi|^2
\\
+\frac{1}2\left(\operatorname{curl}\frac\bP\sigma\right)\cdot(-\Delta^{-1})\left(\operatorname{curl}\frac\bP\sigma\right)
\bigg)\de^3x.
\end{multline*}
The accompanying Poisson bracket functional $\{\mathscr{F},\mathscr{H}\}(\bM,\rho,\bP,\sigma)$ is now a direct sum of two hydrodynamic-type Lie-Poisson brackets, that is, the two-fluid bracket
\begin{multline*}
\{\mathscr{F},\mathscr{H}\}=
{\int}\bM\cdot\bigg(\frac{\delta \mathscr{F}}{\delta \bM}\cdot\nabla\frac{\delta \mathscr{H}}{\delta \bM}-\frac{\delta \mathscr{H}}{\delta \bM}\cdot\nabla\frac{\delta \mathscr{F}}{\delta \bM}\bigg)\de^3 x
-
{\int}\rho\bigg(\frac{\delta \mathscr{F}}{\delta \bM}\cdot\nabla\frac{\delta \mathscr{H}}{\delta \rho}-\frac{\delta \mathscr{H}}{\delta \bM}\cdot\nabla\frac{\delta \mathscr{F}}{\delta \rho}\bigg)\de^3 x
\\
+
{\int}\bP\cdot\bigg(\frac{\delta \mathscr{F}}{\delta \bP}\cdot\nabla\frac{\delta \mathscr{H}}{\delta \bP}-\frac{\delta \mathscr{H}}{\delta \bP}\cdot\nabla\frac{\delta \mathscr{F}}{\delta \bP}\bigg)\de^3 x
-
{\int}\sigma\bigg(\frac{\delta \mathscr{F}}{\delta \bP}\cdot\nabla\frac{\delta \mathscr{H}}{\delta \sigma}-\frac{\delta \mathscr{H}}{\delta \bP}\cdot\nabla\frac{\delta \mathscr{F}}{\delta \sigma}\bigg)\de^3 x.
\end{multline*}
Given the correspondence between Madelung vortohydrodynamics and the Hall MHD system discussed earlier, it is not surprising that the above  bracket is identical to that underlying Hall MHD  \cite{GBRa}. In particular, 
due to the specific form of the Hamiltonian, we have 
$
\sigma {\delta \mathscr{H}}/{\delta \bP}= \hbar \rho {\delta \mathscr{H}}/{\delta \bM}-\hbar {\,\operatorname{curl}^2}\Delta^{-1} ({\bP}/{\sigma})
$,
so that
\[
\frac{\partial\sigma}{\partial t}=-\operatorname{div}\left(\sigma\frac{\delta \mathscr{H}}{\delta \bP}\right)=-\hbar\operatorname{div}\left(\rho\frac{\delta \mathscr{H}}{\delta \bM}\right)
=\hbar\frac{\partial\rho}{\partial t},
\]
and therefore we ensure that the relation $\sigma=\hbar\rho$ is satisfied as a particular solution of
\[
\frac{\partial}{\partial t}(\sigma-\hbar\rho)=0.
\]
Also, using the relations \eqref{chainrule}, we have ${\delta \mathscr{H}}/{\delta \sigma}=\sigma^{-1}\bP\cdot{\delta \mathscr{H}}/{\delta \bP} $, so that the $\bP$-equation reads
\[
\frac{\partial\bP}{\partial t}+\partial_j\bigg( \frac{\delta \mathscr{H}}{\delta {\cal P}_j}\bP\bigg)+{\cal P}_j\nabla \frac{\delta \mathscr{H}}{\delta {\cal P}_j}
=
-\sigma\nabla\left(\frac{1}{\sigma}\frac{\delta \mathscr{H}}{\delta \bP}\cdot\bP\right).
\]
Then, the equation for $\bA=\bP/\sigma$ coincides with \eqref{ohmslaw}, upon setting $\Phi=0$ and $\bv={\delta \mathscr{H}}/{\delta \bP}$. The  vorticity equation \eqref{vortdyn} is obtained by taking the curl, while  \eqref{QFmom} and \eqref{dens} are found from the Hamiltonian equations for  $\bM/\rho$ and $\rho$, respectively. We remark that  the original variable $\bA=\bP/\sigma$ can be made to appear explicitly in the above two-fluid Poisson bracket, thereby recovering the Hall MHD bracket structure in \cite{Holm}.

\section{Disclination dynamics in ideal nematic textures\label{sec:DisclText}}

In this section, we show how the minimal coupling approach  used above to realize vortex dynamics can be extended to the presence of symmetry breaking. In this case, the  group symmetry of the medium is broken by an order parameter  whose isotropy subgroup identifies the residual symmetry in the dynamics. In particular, here we will consider the case of nematic liquid crystals, for which the rotational symmetry is broken by the direction identified by the specific orientation of the nematic molecules.

Before treating the general case of flowing liquid crystals, however, here we apply the previous approach to the orientational dynamics of flowless nematic textures. The molecules in the medium possess a rod-like shape thereby carrying orientational order that evolves over time. Then, the medium is described by the \emph{director field} $\bn(\bx,t)$ such that $|\bn(\bx,t)|=1$. The evolution of this field can be expressed as $\bn(\bx,t)=\chi(\bx,t)\bn_0(\bx)$, where the instantaneous orientational state $\chi(\bx,t)$ is a rotation matrix, that is $\chi(\bx,t)^{-1}=\chi(\bx,t)^{T}$ and $\det \chi(\bx,t)=1$. Generally speaking, disclinations are singularities in the orientational state $\chi(\bx,t)$ although a better characterization needs further details. Before entering the characterization of defects, however, the next section sets up the framework by reviewing nematic texture dynamics in the absence of defects, while the latter will be treated later on.

\subsection{Variational setting of ideal nematic texture dynamics\label{sec:DisclTextSyst}}

As before, here we consider ideal conservative dynamics. While dissipative effects are not treated in this work, we rely on the possibility for those to be included at a later stage.
The conservative equations of motion in the absence of defects may be written as
\begin{equation}\label{NemTextEqn1}
J\rho\frac{d^2\mathbf{n} }{dt^2}-\left(\mathbf{n\cdot h}+J\rho\,\mathbf{n}\cdot\frac{d^2\mathbf{n} }{dt^2}\right)\mathbf{n}+\mathbf{h}=0.
\end{equation}
Here, $J$ is the \emph{microinertia constant} and $\rho$ is the fixed density of the medium, while the \emph{molecular field} 
\[
\mathbf{h}:=\rho\frac{\partial F}{\partial \bn}-\operatorname{div}\bigg(\rho\frac{\partial F}{\partial \nabla\bn}\bigg)
\]
is expressed in terms of the Frank energy. The latter encodes the elastic properties of the medium:
 \begin{align*}
F(\mathbf{n},{\nabla}\mathbf{n})&:=K_2\underbrace{(\mathbf{n}\cdot\operatorname{curl}\mathbf{n})}_{\textit{chirality}}+\frac{1}{2}K_{11}\underbrace{(\operatorname{div}\mathbf{n})^2}_{\textit{splay}}+\frac{1}{2}K_{22}\underbrace{(\mathbf{n}\cdot\operatorname{curl}\mathbf{n})^2}_{\textit{twist}}
+\frac{1}{2}K_{33}\underbrace{|\mathbf{n}\times\operatorname{curl}\mathbf{n}|^2}_{\textit{bend}}.
\end{align*}
As indicated above, each term possesses a precise physical meaning. Equation \eqref{NemTextEqn1} represents a continuum evolution on the unit sphere and represents the basic platform of liquid crystal dynamics.

In this paper, we will resort to an alternative way of writing \eqref{NemTextEqn1} which has the advantage of extracting the instantaneous angular motion; this formulation is due to Volovik \cite{Volovik}. As a first step, we  introduce the angular velocity 
\beq\label{nudef}
\nu_a=\frac12\epsilon_{abc}\hat\nu_{cb}
,\qquad\text{where}\qquad
\hat{\bnu}=\dot{\chi}\chi^{-1}  { \in  \mathfrak{so}(3)} .
\eeq 
Here, $\mathfrak{so}(3)$ denotes the Lie algebra of the rotation group $\mathrm{SO}(3)$. The Lie algebra $\mathfrak{so}(3)$ consists of skew-symmetric $3\times3$ matrices and is isomorphic to the space $\mathbb{R}^3$ of three-dimensional vectors, with the matrix commutator $[\widehat{\bnu},\widehat{\boldsymbol{w}}]$ corresponding to the cross product $\bnu\times\boldsymbol{w}$. More precisely, this isomorphism between $(\mathfrak{so}(3),[\,,])$ and $(\Bbb{R}^3,\times)$ is provided by the map $\bnu\mapsto\widehat{\bnu}$, given by $\hat\nu_{ab}=-\epsilon_{abc}\nu_{c}$, so that $\hat{\bnu}\boldsymbol{w}:= \bnu \times \boldsymbol{w}$ for any two arbitrary vectors $\bnu$ and $\boldsymbol{w}$.

Given the angular velocity in \eqref{nudef}, equation \eqref{NemTextEqn1} can be conveniently written in terms of the angular momentum $\bmu=J\rho\bnu$ as
\beq\label{texteqs}
\frac{\partial\bmu}{\partial t}=-\bn\times\bh
,\qquad\qquad
\frac{\partial\bn}{\partial t}=\frac1{J\rho}\bmu\times\bn.
\eeq
Here, $\rho$ is the fixed density of the medium and we observe the conservation law
\beq\label{Noether1}
\frac{\partial}{\partial t}(\bmu\cdot\bn)=0,
\eeq
which is the direct consequence of the texture's rotational invariance around the director axis, so that the molecular spin 
\beq\label{parmom}
\mu_\|=\bmu\cdot\bn
\eeq
around $\bn$ is an irrelevant Noether charge.  This $SO(2)$-invariance is indeed the essence of symmetry breaking: at the single molecule level, the full $SO(3)$-symmetry is `broken' thereby leaving the $SO(2)$-symmetry as a leftover, so that the dynamics  takes place on the director  space $SO(3)/SO(2)\sim S^2$, that is the unit sphere. Equation \eqref{NemTextEqn1} is derived from \eqref{texteqs} by setting $\bmu\cdot\bn=0$, after evaluating $\bn\times(\bn\times\partial^2_t\bn)$.

The texture equations \eqref{texteqs} can be obtained from the following variational principle:
\beq\label{nemtextvarprin}
\delta{\int_{t_1}^{t_2}}{\int}\bigg(
\bmu\cdot\bnu-\frac12\frac{|\bmu|^2}{J\rho}-\rho F(\bn,\nabla\bn)
\bigg)\de^3x\de t=0,
\eeq
with respect to arbitrary variations $\delta\bmu$ and variations $\delta\bnu$ and $\delta\bn$ induced by arbitrary variation $\delta\chi$. Upon denoting 
 $\widehat{\bxi}=(\delta\chi)\chi^{-1} \in  \mathfrak{so}(3)$, one has
\beq\label{nemvars}
\delta\bnu=\partial_t\bxi+\bxi\times\bnu
,\qquad\qquad
\delta\bn=\bxi\times\bn.
\eeq
In addition, the evolution law $\bn=\chi\bn_0$ leads to the evolution equation 
\beq\label{auxneq}
\partial_t\bn=\bnu\times\bn
\eeq
as in standard Euler-Poincar\'e theory \cite{HoMaRa98}.

\subsection{Wryness tensor, disclination density, and its vector potential\label{sec:discldens}}

The rotational strain of a nematic texture is identified with the \emph{wryness tensor} 
\beq\label{gammadef}
\gamma_{aj}=-\frac12\epsilon_{abc}\hat\gamma_{bcj}
,\qquad\text{where}\qquad
\widehat{\bgamma}=\chi(\nabla\chi^{-1})+
\chi\widehat{\bgamma}_0\chi^{-1}.
\eeq
Apart from the initial value $\hat\gamma_0$ and a slightly different choice of convention, this quantity appears prominently in Kl\'eman's early work on liquid crystals \cite{Kleman}. As we will see, the initial value is absolutely crucial in defect dynamics. The  role of the wryness tensor as a  gauge vector potential, established in \cite{GBRa}, is reflected in the parallel transport relation
\begin{equation}\label{gammaEvol}
\partial_j \bn=\bn\times \bgamma_j,
\end{equation}
which is found by simply taking the gradient of the evolution law $\bn =\chi \bn_0 $ and writing $\partial_j\bn_0 =\bn_0\times\bgamma_{0j}$. Notice that the latter can be inverted to write
\beq\label{wrynessdef}
\bgamma_j=\partial_j\bn\times\bn+\gamma_{{\scriptscriptstyle\|}j}\bn,
\eeq
where
\[
\bgammap=n_a\bgamma_a.
\]
We emphasize that we use $a,b,c,\dots$ as vector indices and $i,j,k,\dots$ as spatial coordinate indices.

In his micropolar theory of liquid crystals \cite{Eringen}, Eringen discarded $\bgammap$ assuming it could be made to vanish at all times. As pointed out by the authors in \cite{GBRaTr2012,GBRaTr2013}, even if one chooses $\bgamma_{{\scriptscriptstyle\|}0}=0$ at the initial time, the dynamics develop nonzero values of $\bgammap$ that cannot be ignored. Upon recalling $\bnu=J^{-1}\bmu/\rho$, this is easily shown by looking at the equation
\beq\label{gammaevol}
\frac{\partial\bgamma_j}{\partial t}+\frac{\partial\bnu}{\partial x^j}+\bgamma_j\times\bnu=0,
\eeq
following from \eqref{gammadef} so that, upon denoting $\nu_{\scriptscriptstyle\|}=\bnu\cdot\bn$,
\[
\frac{\partial\gamma_{{\scriptscriptstyle\|}j}}{\partial t}=-\frac{\partial\nu_{\scriptscriptstyle\|}}{\partial x^j}-\bn\cdot\bnu\times \bgamma_j.
\]
We observe that the right-hand side above cannot be made to vanish in general.

Despite its intricacies, the wryness tensor becomes crucial in providing the definition of the disclination density, although several variants have been reported in the literature. For example, Volovik \cite{Volovik} identified the disclination density with the curvature field associated to $\bgamma$, 
that is,
\beq\label{curv1}
\bOmega_j = \frac12\epsilon_{ijk}\widehat{\bOmega}_{jk}
,\qquad\text{where}\qquad
\widehat{\bOmega}_{ij}=\partial_i\bgamma_j-\partial_j\bgamma_i+\bgamma_i\times \bgamma_j.
\eeq
Similarly to Volovick, Kl\'eman \cite{Kleman} wrote that ``the relation $\Omega_{aj}=0$ means that the body does not contain any disclination'' (quote). However, Halperin \cite{Halperin} wrote a disclination line exactly in the vortex form  \eqref{vortex}, thereby indicating a discrepancy in the tensorial nature of the different objects.  Another version identifies the disclination density with 
\begin{equation}\label{Mermin_Ho}
\operatorname{curl}\bv_\MH=\frac12\epsilon_{abc}n_a\nabla n_b\times \nabla n_c
\end{equation}
where $\bv_\MH$ is the Mermin-Ho velocity, determined by this relation only up to the addition of an irrotational vector field, that is a pure a gradient term.

We will now show that, perhaps not surprisingly, these relations are all related via the parallel transport relation \eqref{gammaEvol}. Indeed, as shown in \cite{FoTr}, taking the curl of the latter yields
\beq\label{OmPerp}
\bOmega_j\times\bn=0
\qquad\implies\qquad
\bOmega_j=\Omega_{\|j} \bn.
\eeq
Thus, the only nontrivial component of the curvature field that allows for  \eqref{gammaEvol} to hold is the component parallel to the director. This result seems to have gone unnoticed in the literature. While this sheds  light on why different authors looked at quantities with different dimensions, the question remains whether $\bOmega_\|$
generally inherits any of the geometric features enjoyed by $\bOmega$ itself. To answer this question, we use \eqref{gammaEvol} to compute
\begin{align*}
\partial_i\gamma_{{\scriptscriptstyle\|}j}-\partial_j\gamma_{{\scriptscriptstyle\|}i}=&\,\partial_i(\bn\cdot\bgamma_j)-\partial_j(\bn\cdot\bgamma_i)
\\
=&\,
(\bn\times\bgamma_i)\cdot\bgamma_j
+
\bn\cdot\partial_i\bgamma_j
-
(\bn\times\bgamma_j)\cdot\bgamma_i
-
\bn\cdot\partial_j\bgamma_i
\\
=&\,
\bn\cdot\widehat{\bOmega}_{ij}
+
\bn\cdot(\bgamma_i\times\bgamma_j)
,
\end{align*}
where the last equality follows from \eqref{curv1}.
Since the second term depends on $\bgamma$ only through $\bgamma_\perp=\nabla \bn\times \bn$, we have
\begin{align*}
\bn\cdot(\bgamma_i\times\bgamma_j)&=
\bn\cdot\big((\partial_i \bn\times \bn)\times(\partial_j \bn\times \bn)\big)
\\
&=
\bn\cdot\partial_i\bn\times\partial_j \bn
,
\end{align*}
giving
\[
\bn\cdot\widehat{\bOmega}_{ij}=\partial_i\gamma_{{\scriptscriptstyle\|}j}-\partial_j\gamma_{{\scriptscriptstyle\|}i}-\bn\cdot\partial_i\bn\times\partial_j \bn.
\]
Multiplying this relation by $\epsilon_{kij}$, using $(\operatorname{curl} w)_k=\epsilon_{kij}(\partial_iw_j-\partial_jw_i)/2$ as well as \eqref{Mermin_Ho} yields, 
\beq
{\bOmega_\|}=\operatorname{curl} (\bgammap-\bv_\MH).
\eeq
Thus, if the quantity
\beq\label{connform}
\bA=\bgammap-\bv_\MH
\eeq
 identifies a vector potential for an $SO(2)$ gauge, then $\bOmega_\|=\operatorname{curl}\bA$ is its associated curvature field. To show that  $\bA$ does transform as a $SO(2)$-gauge potential, we notice that \eqref{gammaevol} leads to the equation of motion \cite{GBRa,GBRaTr2013,Holm2002}
\[
\frac{\partial\bOmega_j}{\partial t}=\bnu\times\bOmega_j,
\]
and thus, using $\partial_t\bn=\bnu\times\bn$,
\[
\frac{\partial \bOmega_\|}{\partial t}=0.
\]
In turn, this means that there is a function $\Phi(\bx,t)$ such that
\[
\frac{\partial\bA}{\partial t}+\nabla\Phi=0,
\]
and therefore the quantity \eqref{connform} indeed transforms as a vector potential for an $SO(2)$ gauge. 

In the light of the above, we are led to refer to
\beq\label{discldens}
\bomega:=\bOmega_\|=\operatorname{curl}\bA
,
\eeq
as the \emph{disclination density},
while $\bA$ will be called the \emph{disclination potential}.
We emphasize that this definition is indeed equivalent to Volovik's in \eqref{curv1}, due to \eqref{OmPerp}. We stress, however, that $\bgamma$ and $\bA$ are $SO(3)$ and $SO(2)$ gauge potentials, respectively, and they should be treated as distinctively different quantities. This fact clarifies why the dimensionality of the different objects considered in the literature differs depending on the author. Nevertheless, we remark that, in general,  $\bgammap\neq0$, and thus $\bomega\neq\operatorname{curl}\bv_\MH$, which contradicts some of the recent works in the field. The relation $\bgamma_{{\scriptscriptstyle\|}0}=0$ represents a natural initial condition, so that $\bomega_{0}=-\operatorname{curl}\bv_{\MH0}$ holds at the initial time only. Here, the minus sign is a result of our sign convention in defining the wryness tensor \eqref{gammadef}.

\subsection{Interaction dynamics of disclinations and texture fields\label{sec:coupleddisclnem}}

Once the geometric setting underlying the disclination density has been developed, we are now ready to extend the construction from Section \ref{sec:quantvort} to the present case of ideal nematic textures. In proceeding analogously, we modify the variational principle \eqref{nemtextvarprin} as follows:
\begin{multline}\label{nemtextvarprin2}
\delta{\int_{t_1}^{t_2}}{\int}\bigg(
\bmu\cdot(\bnu+\Phi\bn)-\frac12\frac{|\bmu|^2}{J\rho}-\rho F(\bn,\nabla\bn)
-
\bA\cdot\operatorname{curl}\bv_\MH
\\
+\sigma (\bA\cdot \bv-\Phi)-\frac{1}2\bomega\cdot(-\Delta^{-1})\bomega
\bigg)\de^3x\de t=0,
\end{multline}
where the variables $\bA$ and $\bn$ are varied independently, while $\operatorname{curl}\bv_\MH$ is expressed in terms of $\bn$ by \eqref{Mermin_Ho} and $\bomega=\operatorname{curl}\bA$. Indeed, $\delta\bmu$, $\delta\bA$, and $\delta\Phi$ are arbitrary, while $\delta\bnu$ and $\delta\bn$ are  given in \eqref{nemvars}, and $\delta\bv$ and $\delta\sigma$  in \eqref{vars1}. In addition, we have the auxiliary equations \eqref{GCT} and \eqref{auxneq}.

In the action identified by \eqref{nemtextvarprin2}, the first $\Phi$-term is a temporal gauge term, while the second line is identical to the second line of \eqref{VarPr1}. While the last term is manifestly gauge invariant, the gauge invariance of the first term can be understood by taking arbitrary variations $\delta\bmu$, which gives
\begin{equation}\label{Legendre_T}
\bmu= J\rho( \bnu + \Phi\bn).
\end{equation}
Then, leaving the momentum variable $\bmu$ unaffected and transforming $\Phi\mapsto\Phi-\partial_t\varphi$ yields the following gauge transformation rule for $\bnu$:
\beq\label{gaugenu}
\bnu\mapsto\bnu +\bn \partial_t\varphi.
\eeq
With this transformation rule, the first term in \eqref{nemtextvarprin2} is manifestly gauge invariant.

The last term in the first line of \eqref{nemtextvarprin2} is a new term that deserves a certain attention. This term is inserted here to realize the coupling between the texture and the disclinations. Its expression is based on two features: 1) the tensorial nature of the disclination potential and of $\operatorname{curl}\bv_\MH$ and 2) the absence of a unique gauge-independent expression for $\bv_\MH$. In pursuing point 1, we see that both $\bA$ and $\bv_\MH$ are differential one forms. Then, an interaction energy must be given in such a way that the combination of these two objects identifies an integration measure. Notice that, being a differential one-form, the Mermin-Ho velocity  $v_\MH=\bv_\MH\cdot\de\bx$ can  hardly be treated as a hydrodynamic velocity vector field in the absence of a metric. In the present case, the integration measure in 3D space is constructed as $\omega\wedge v_\MH=\bomega\cdot\bv_{\MH\,}\de^3 x$. Then, integration by parts $\int \bomega\cdot\bv_{\MH\,}\de^3 x=\int\bA\cdot\operatorname{curl}\bv_{\MH\,}\de^3 x$ allows to address point 2, thereby overcoming the gauge freedom in the definition of $\bv_{\MH\,}$.
One may also think of replacing the interaction energy $\int\bomega\cdot\bv_\MH\,\de^3 x$ by the entire disclination helicity  
\beq\label{disclhel}
{\cal H}_{\rm d}(\bA)={\int}\bomega\cdot\bA\,\de^3 x.
\eeq
However, as we will see, this quantity is a dynamical invariant, similarly to the vortex helicity in \eqref{vorthel}, and therefore cannot be used to realize an interaction energy.

Arbitrary variations $\delta\Phi$ and  $\delta\bA$ yield
\beq\label{charge&vel}
\sigma=\mu_\|
\,,\qquad\qquad
\bv=\frac1\sigma\operatorname{curl}(\bv_\MH-\Delta^{-1}\bomega)
,
\eeq
where we recall the definition \eqref{parmom}.
In order to find the remaining equations, we need to evaluate
\beq\label{varder}
\frac{\delta}{\delta\bn}
{\int}\bA\cdot\operatorname{curl}\bv_\MH\,\de^3x 
=
\frac32\epsilon_{ijk}{\cal A}_i \partial_j\bn\times\partial_k \bn
-
    \bn\times(\bomega\cdot\nabla) \bn.
\eeq
The Euler-Poincar\'e variations $\delta\bnu$ and $\delta\bn$ then yield, in place of equations \eqref{texteqs},
\beq\label{texteqsdef}
\frac{\partial\bmu}{\partial t}=-\bn\times\bh
-(\bomega\cdot\nabla) \bn
,\qquad\qquad
\frac{\partial\bn}{\partial t}=\frac1{J\rho}\bmu\times\bn,
\eeq
where we used that the first term in \eqref{varder} is parallel to $\bn$.
We observe the explicit appearance of the disclination density in the angular momentum equation, thereby realizing the texture-disclination coupling. We emphasize that the equations above  do not affect the gauge charge conservation \eqref{Noether1}.
Finally, the Euler-Poincar\'e variations $\delta\bv$ and $\delta\sigma$ yield the disclination potential evolution:
\beq\label{disclpotevol}
\frac{\partial \bA}{\partial t}+\nabla\Phi=\bv\times\bomega,
\eeq
where, from \eqref{charge&vel}, $\bv$ is given as
\beq\label{disclvel}
\bv=\frac1{\mu_\|}\operatorname{curl}(\bv_\MH-\Delta^{-1}\bomega)
.
\eeq
The advection equation $\partial_t\sigma+ \operatorname{div}(\sigma \bv)=0$ is trivially satisfied since $\partial_t\sigma = \partial_t \mu_\|=0$ by \eqref{texteqsdef} and $\operatorname{div}(\sigma\bv)=0$ by \eqref{disclvel}. Notice that equation \eqref{disclpotevol}, toghether with the relation \eqref{discldens}, shows that the disclination helicity ${\cal H}_{\rm d}$ in \eqref{disclhel} is conserved by the dynamics.

In conclusion, upon taking the curl of \eqref{disclpotevol}, the final system consists of the equations 
\begin{align}\label{mudynwdiscl}
\frac{\partial\bmu}{\partial t}=&\,-\rho\bn\times\bh
-(\bomega\cdot\nabla) \bn,
\\\label{ndynwdiscl}
\frac{\partial\bn}{\partial t}=&\,\frac1{J\rho}\bmu\times\bn,
\\\label{disclindenseqn}
\frac{\partial \bomega}{\partial t}=&\,{\operatorname{curl}}\bigg(\frac1{\bmu\cdot\bn}\operatorname{curl}(\bv_\MH-\Delta^{-1}\bomega)\times\bomega\bigg).
\end{align}
We observe that a nontrivial disclination density dynamics enforces a nonzero gauge charge $\mu_\|\neq0$. This requirement  means that the gauge charge is intrinsically related to the disclination geometry. This fact  explains why the Ericksen-Leslie model, associated with equation \eqref{NemTextEqn1} in the presence of a hydrodynamic liquid crystal flow, is insufficient to capture disclinations. Indeed, as noted above, equation \eqref{NemTextEqn1} requires $\mu_\|=0$, which is incompatible with the presence of a disclination density. However, as we have just shown, Volovik's formulation in equations \eqref{texteqs} can be extended to include disclination dynamics. 

\subsection{Lagrangian and Hamiltonian formulations\label{sec:HamDisclNem}}

Similarly to Madelung vortohydrodynamics, the system \eqref{mudynwdiscl}-\eqref{ndynwdiscl} possesses a Lagrangian formulation that is the analogue of the variational principle \eqref{MadVHD-vp}, although in the absence of hydrodynamic flow. Such a Lagrangian formulation involves only velocity variables, in this case the angular velocity \eqref{nudef}. This formulation may be found by replacing the Legendre transform relation \eqref{Legendre_T}
in \eqref{nemtextvarprin2}, thereby obtaining Hamilton's variational principle $\delta\int_{t_1}^{t_2}\!\ell\de t=0$, with the Lagrangian functional $\ell=\ell(\bnu,\bn,\bA,\,\,\,\dot{\!\!\!\boldsymbol{\cal A}},\Phi,\dot\Phi)$ given by
\beq\label{textLagr}
\ell={\int}\bigg(
\frac{J}2\rho|\bnu+\Phi\bn|^2+\sigma (\bA\cdot \bv-\Phi)-\rho F(\bn,\nabla\bn)
-
\bA\cdot\operatorname{curl}\bv_\MH
\\
-\frac{1}2(\operatorname{curl}\bA)\cdot(-\Delta^{-1})(\operatorname{curl}\bA)
\bigg)\de^3x.
\eeq
Here, the variations are the same as in the previous section and we recall that $\operatorname{curl}\bv_\MH$ is entirely expressed in terms of the director $\bn$ and its gradient. Notice that choosing the temporal gauge $\Phi=0$ and setting $\bA=0$  recovers the Euler-Poincar\'e Lagrangian $\ell=\ell(\bnu,\bn)$ which underlies the equations \eqref{texteqs} holding in the absence of disclinations \cite{GBTr2010,GBRaTr2012,GBRaTr2013}.

Having provided the Lagrangian formulation, here we also present the Hamiltonian structure of the system \eqref{mudynwdiscl}-\eqref{ndynwdiscl}. As usual, we start by identifying the Hamiltonian functional with the total energy
\beq\label{text-dinen}
\mathscr{H}(\bmu, \bn,
\bomega)={\int}\left(\frac1{2J}\frac{|\bmu|^2}{\rho}+\rho F(\bn,\nabla\bn)
+
\bomega\cdot\bv_\MH
+\frac{1}2\bomega(-\Delta^{-1})\bomega\right)\de^3 x.
\eeq
While the Memerin-Ho velocity $\bv_\MH$ appearing in the third term does not have a unique expression in terms of the director $\bn$, one may overcome this gauge freedom as before by rearranging $\int\bomega\cdot\bv_\MH\de^3x=\int(\operatorname{curl}^{-1}\bomega)\cdot \operatorname{curl}\bv_\MH\de^3x=-\int(\operatorname{curl}\Delta^{-1}\bomega)\cdot\operatorname{curl}\bv_\MH\de^3x$,
thereby obtaining a closed form of the texture-disclination interaction energy. 
In order to complete the derivation of the noncanonical Hamiltonian structure governing \eqref{mudynwdiscl}-\eqref{ndynwdiscl}, we shall proceed as in \S\ref{sec:MCHDHamSt} and write the original variational principle \eqref{nemtextvarprin2} for a general Hamiltonian functional $\mathscr{H}$ as
\[
\delta{\int_{t_1}^{t_2}}{\int}\bigg(
\bmu\cdot(\bnu+\Phi\bn)+\sigma (\bA\cdot \bv-\Phi) -\mathscr{H}(\bmu,\bn,\bomega)
\bigg)\de^3x\de t=0,
\]
which in the present context plays the role of \eqref{DF_H}.
Here, the arbitrary variations $\delta\bmu$, $\delta \Phi$, and $\delta\bA$ yield, respectively,
\beq\label{vareqns1}
\bnu+ \Phi \bn = \frac{\delta \mathscr{H}}{\delta\bmu},\qquad\quad \sigma=\mu_\| , \qquad\quad \sigma\bv = \operatorname{curl} \frac{\delta \mathscr{H}}{\delta\bomega},
\eeq
while the Euler-Poincar\'e variations $(\delta \bnu, \delta\bn)$, and $(\delta \bv,\delta \sigma)$ lead to, respectively,
\beq\label{vareqns2}
\frac{\partial \bmu}{\partial t} + \bmu \times \frac{\delta \mathscr{H}}{\delta\bmu}=-\bn\times\frac{\delta \mathscr{H}}{\delta\bn} , \qquad \frac{\partial \bn}{\partial t} = \bnu\times\bn,
\qquad
\frac{\partial \bA}{\partial t}+ \nabla\Phi = \bv\times\operatorname{curl}\bA, \qquad \frac{\partial \sigma}{\partial t} + \operatorname{div}(\sigma \bv)=0,
\eeq
where we have included the auxiliary equations \eqref{GCT} and \eqref{auxneq}.
Using the  relations \eqref{vareqns1} in  \eqref{vareqns2} an taking the curl of the $\bA$-equation, we get
\[
\frac{\partial \bmu}{\partial t}  + \bmu \times  \frac{\delta \mathscr{H}}{\delta\bmu} =-\bn\times\frac{\delta \mathscr{H}}{\delta\bn}, \qquad \frac{\partial \bn}{\partial t}  = \frac{\delta \mathscr{H}}{\delta\bmu}\times\bn,
\qquad
\frac{\partial \bomega}{\partial t}  ={\operatorname{curl}}\bigg( \frac{1}{\sigma} \operatorname{curl} \frac{\delta \mathscr{H}}{\delta\bomega} \times \bomega\bigg), \qquad \frac{\partial \sigma}{\partial t} =0.
\]
From the last equation, we recover the constant $\sigma=\mu_\|= \bmu\cdot\bn$ previously identified in \eqref{Noether1}  using the first two equations above. From the general form of the first three equations  for an arbitrary Hamiltonian functional $\mathscr{H}$, it is now apparent that they are Hamiltonian with respect to the Poisson bracket
\begin{multline}\label{texturebkt}
\{\mathscr{F},\mathscr{H}\}(\bmu,\bn,\bomega)={\int}\bigg[\bmu\cdot\frac{\delta \mathscr{F}}{\delta \bmu}\times\frac{\delta \mathscr{H}}{\delta \bmu}+\bn\cdot\bigg(\frac{\delta \mathscr{F}}{\delta \bn}\times\frac{\delta \mathscr{H}}{\delta \bmu}-\frac{\delta \mathscr{H}}{\delta \bn}\times\frac{\delta \mathscr{F}}{\delta \bmu}\bigg)\bigg]\de^3 x
\\
+{\int}
\frac{\bomega}{\bmu\cdot\bn}\cdot\left(\operatorname{curl}\frac{\delta \mathscr{F}}{\delta \bomega}\right)\times\left(\operatorname{curl}\frac{\delta \mathscr{H}}{\delta \bomega}\right)
\de^3 x.
\end{multline}
\color{black}

Similarly to \S\ref{sec:MCHDHamSt}, the above Poisson bracket is the direct sum of two Lie-Poisson brackets: the first line is the Lie-Poisson bracket governing nematic textures in the absence of disclinations \cite{GBRa,GBTr2010,Holm2002,Volovik}, the second line is again a Lie-Poisson bracket of hydrodynamic type. However, unlike those encountered in \S\ref{sec:MCHDHamSt}, the second line above applies to a hydrodynamic flow preserving the volume form $\mu_\|(\bx)\de^3x$ identified by  $\mu_{\|}= \bmu\cdot \bn$.
In the case $\mu_\|=1$, the disclination flow is incompressible and a further reduction may be considered in terms of vortex filaments via the Rasetti-Regge structure for superfluid helium-II \cite{Holm03,RasettiRegge,PennaSpera1989}. Instead of pursuing this direction, the next section introduces a regularization scheme for the interaction of a disclination line and the surrounding texture.

\subsection{Regularized disclination line motion and Rasetti-Regge bracket\label{sec:BS-RR}}

In this section, we  use the Biot-Savart law to study disclination lines in the simple case  $\mu_\|=1$. Following  Halperin's work \cite{Halperin}, we introduce a disclination filament by writing \cite{Schimming2}
\beq\label{disclinationline}
\bomega(\bx, t)=n\pi {\int}\bR_s(s,t)\delta(\bx-\bR(s,t))\de s,
\eeq
so that we use Coulomb's gauge ${\operatorname{div}}\bA=0$ to write $\bA=\operatorname{curl}^{-1}\bomega$, thereby recovering \eqref{A_t} with
$\bE= n\pi  \int\bR_s\times {\bR}_t\,\delta(\bx-\bR)\,\de s$.
As in  Madelung vortohydrodynamics, this case leads to common divergences in the theory of defect dynamics. 

To proceed further, here we adopt the regularization already proposed in \S\ref{sec:vortlines}. In particular, we replace the action principle $\delta\int_{t_1}^{t_2}\ell\,\de t=0$ associated to the Lagrangian \eqref{textLagr} by
\[
\delta{\int_{t_1}^{t_2}}{\int}\bigg(
\bmu\cdot(\bnu+\bar\Phi\bn)+\sigma (\bA\cdot \bv-\Phi) -\frac1{2J}\frac{|\bmu|^2}{\rho}-\rho F(\bn,\nabla\bn)
-
{\bar\bomega}\cdot\bv_\MH
-\frac{1}2\bar\bomega(-\Delta^{-1})\bar\bomega
\bigg)\de^3x\de t=0,
\]
where $\bar\bomega=\operatorname{curl}\barA$ and  $(\bar\Phi,\barA)=(K*\Phi,K*\bA)$ are the result of a filtering process using a translation-invariant mollifier $K(\bx,\bx')=K(\bx-\bx')$  such that $K(\bx)= K(-\bx)$ and ${\int} K(\bx-\bx')\,\de^3 x=1$. In analogy with \eqref{reggauge}, the above action principle is invariant under the regularized gauge transformation 
\beq\label{reggauge2}
\bA\mapsto \bA+\nabla \varphi
,\qquad\qquad
\Phi\mapsto \Phi-\partial_t \varphi
,\qquad\qquad
\bnu\mapsto\bnu +\bn \partial_t\bar\varphi,
\eeq
holding for a fixed convolution filter $K$. Then, taking variations yields
\beq\label{charge&velbar}
\sigma=\overline\mu_\|
\,,\qquad\qquad
\bv=\frac1\sigma\operatorname{curl}\overline{(\bv_\MH-\Delta^{-1}\bar\bomega)}
,
\eeq
as well as
\beq\label{vareqns2bar}
\frac{\partial\bmu}{\partial t}=-\rho\bn\times\bh
-(\bar\bomega\cdot\nabla) \bn,
\qquad\quad
\frac{\partial\bn}{\partial t}=\frac\bmu{J\rho}\times\bn,
\eeq
while \eqref{disclpotevol} remains unaffected. Notice that, in analogy with \eqref{varder}, the first equation in \eqref{vareqns2bar} follows from
$ 
\delta
\int\bar\bomega\cdot\bv_\MH\,\de^3x 
=\int\delta\barA\cdot\operatorname{curl}\bv_\MH\,\de^3x 
+
\int \delta\bn\cdot(3\epsilon_{ijk}
\bar{\!{\cal A}\,}\!_i \partial_j\bn\times\partial_k \bn
-
  \bn\times(\bar\bomega\cdot\nabla) \bn)\,\de^3x
$, while the regularized disclination velocity $\bv$ preserves the gauge  charge density \eqref{parmom}, so that
\[
\partial_t\sigma=-{\operatorname{div}}(\sigma\bv)=0=\partial_t\overline{\mu_\|},
\]
where the last equality follows from the incompressibility relation $\partial_t{\mu_\|}=0$ obtained from \eqref{vareqns2bar}.  Here, the first condition in \eqref{charge&velbar} is preserved as the  Noether quantity associated to the invariance under the regularized gauge transformations \eqref{reggauge2}.

At this point, one introduces the disclination filament \eqref{disclinationline} and  resorts to the Coulomb gauge so that  ${\operatorname{div}}\barA=\overline{{\operatorname{div}}\bA}=0$ and $\barA=\operatorname{curl}^{-1}\bar\bomega$. Inserting  \eqref{A_t} in ${\partial_t \bA}+\nabla \Phi={\bv}\times \operatorname{curl}\bA$ gives the ideal Ohm's law  $\bE+\bv\times\bomega=0$, thereby leading to 
\beq\label{MagnusDiscl}
\bR_s\times({\bR}_t-{\bv}(\bR))=0,
\qquad\text{where}\qquad
\bv(\bR)=\left.\frac1{\overline{\mu_\|}}\operatorname{curl}\overline{(\bv_\MH-\Delta^{-1}\bar\bomega)}\right|_{\bx=\bR}
\eeq
and $\overline{\mu_\|}$ is a constant density. Then, together with the regularized texture equations \eqref{vareqns2bar}, the disclination equation above comprises
a regularized \emph{Biot-Savart nematodynamics} model governing the time evolution of $\bmu(\bx, t)$, $\bn(\bx, t)$, and $\bR(s,t)$ in  ideal nematic textures.  Notice that the  Biot-Savart energy term in the disclination velocity $\bv(\bR)$ corresponds to the  Peach-Koehler  force in defect dynamics \cite{Carpio,Peach}. 

The proposed regularized model of Biot-Savart nematodynamics possesses the following variational formulation:
\begin{multline}\label{nemtextvarprin3}
\delta{\int_{t_1}^{t_2}}\bigg[\frac{\pi p}3{\int}\bR_t\cdot\bR\times\bR_s\de s+{\int}\bigg(
\bmu\cdot\bnu
-\frac12\frac{|\bmu|^2}{J\rho}-\rho F(\bn,\nabla\bn)
\\
-
{\operatorname{curl}}^{-1}\bar\bomega \cdot {\operatorname{curl}} \bv _\MH - \frac{1}{2} \bar\bomega  \cdot(-\Delta^{-1})\bar\bomega\bigg)\de^3x\bigg]
\de t=0,
\end{multline}
where we have set $\mu_\|=1=\overline{\mu_\|}$ for simplicity, and 
\[
\bar\bomega(\bx,t)=n\pi {\int}\bR_s(s, t) K(\bx-\bR(s,t))\de s
\]
is a functional of the disclination line $\bR(s,t)$.
We observe the appearance in the first  term of \eqref{nemtextvarprin3} of the Rasetti-Regge elementary action for vortex filaments \cite{RasettiRegge}, which governs  vortex dynamics in  incompressible flows.  
The presence of the Rasetti-Regge action in the above action principle determines a peculiar form of the Hamiltonian structure underlying the coupled evolution of disclination lines and background fields in nematic textures. In particular, the first term in \eqref{nemtextvarprin3}  is responsible for the appearance of the Rasetti-Regge bracket structure \cite{RasettiRegge}
\beq\label{RRbkt}
\{{F},{H}\}_{\RR}(\bR)=\frac1{\pi p}{\oint}\frac{\bR_s}{|\bR_s|^2}\cdot\frac{\delta {F}}{\delta \bR}\times\frac{\delta {H}}{\delta \bR}\,\de s
\eeq
generally governing  vortex dynamics in superfluid helium-II. In the present case, the Hamiltonian structure of the system comprised by \eqref{charge&velbar}-\eqref{vareqns2bar} is given by the Poisson bracket 
\begin{equation}\label{textureRRbkt}
\{ \mathscr{F}, \mathscr{H}\}(\bmu,\bn,\bR)=\{ \mathscr{F}, \mathscr{H}\}_{\RR}+{\int}\bigg[\bmu\cdot\frac{\delta \mathscr{F}}{\delta \bmu}\times\frac{\delta \mathscr{H}}{\delta \bmu}+\bn\cdot\bigg(\frac{\delta \mathscr{F}}{\delta \bn}\times\frac{\delta \mathscr{H}}{\delta \bmu}-\frac{\delta \mathscr{H}}{\delta \bn}\times\frac{\delta \mathscr{F}}{\delta \bmu}\bigg)\bigg]\de^3 x
\end{equation}
and the Hamiltonian functional
\beq
\mathscr{H}(\bmu, \bn,\bR)={\int}\bigg(
\frac12\frac{|\bmu|^2}{J\rho}+\rho F(\bn,\nabla\bn)
+
{\operatorname{curl}}^{-1}\bar\bomega \cdot {\operatorname{curl}} \bv _\MH + \frac{1}{2} \bar\bomega  \cdot(-\Delta^{-1})\bar\bomega\bigg)\de^3 x.
\eeq
To our knowledge, it is the first time that the Rasetti-Regge bracket structure appears in the context of nematic media. We also point out that the present case of spatially constant gauge charge $\mu_\|=1$ can  be extended to the case  of inhomogenous media by replacing  $\bnu\rightsquigarrow\bnu-\bn\operatorname{div}\Delta^{-1}\bar\bE$ in \eqref{nemtextvarprin3} and resorting to suitable generalizations of the Rasetti-Regge construction, as in \cite{Ruban}.

The Rasetti-Regge bracket \eqref{RRbkt} involves a rich geometric content which is intrinsic in the dynamics of vortex filaments, as outlined in \cite{Holm03,MaWe83,PennaSpera1989} within the context of reduction by symmetry. The extension of vortex geometry to the present case involving nematic texture dynamics with disclinations represents a most interesting direction that is left for a separate venue. In what follows, we will see how the previous discussion adapts to the more general case of flowing liquid crystals with disclinations.


\section{Disclination dynamics in ideal liquid crystal flows\label{sec:liqxtals}}
Having formulated the coupled dynamics of disclinations and floweless nematic textures in the preceding sections, here we move on to introduce the hydrodynamic flow. As mentioned before,   the standard Ericksen-Leslie theory of liquid crystals is  insufficient for capturing disclination dynamics. In our setting, we saw that this is due to the fact that the Ericksen-Leslie theory assumes $\mu_\|=\bmu\cdot\bn=0$ throughout, which is hardly compatible with the presence of disclinations in the system. Therefore, here we will stick to Volovik's formulation in such a way to treat the angular momentum $\bmu$ as a standalone dynamical variable.

\subsection{Variational principle and equations of motion\label{sec:LiqxtalEqns}}
The presence of the hydrodynamic flow is readily incorporated in the previous formulation by adding the fluid energy terms (kinetic and internal) to the previous variational principle \eqref{nemtextvarprin2}, which leads to
\begin{multline}\label{ELDisclvarprin}
\delta{\int_{t_1}^{t_2}}{\int}\bigg(\frac{m}2\rho{|\bu|^2}-\rho{U}(\rho)+
\bmu\cdot\big(\bnu+(\Phi-\bA\cdot\bu)\bn\big)-\frac12\frac{|\bmu|^2}{J\rho}-\rho F(\bn,\nabla\bn)
\\
-
\bA\cdot\operatorname{curl}\bv_\MH
-\frac{1}2\bomega\cdot(-\Delta^{-1})\bomega+\sigma (\bA\cdot \bv-\Phi)
\bigg)\de^3x\de t=0.
\end{multline}\noindent
Besides the internal fluid energy in the second term of the first line, we also notice that the temporal gauge term in the first line has been modified from \eqref{nemtextvarprin2} to reflect the minimal coupling procedure in the presence of a velocity vector field, that is, we have replaced $\Phi \rightsquigarrow \Phi-\bA\cdot\bu$ in \eqref{nemtextvarprin2}.
The variations in \eqref{ELDisclvarprin} are taken as follows:  $\delta\bmu$, $\delta\Phi$, and $\delta\bA$ are arbitrary, $\delta\bu$ and $\delta\rho$ are expressed exactly as in \eqref{vars2}, while $\delta\bv$ and $\delta\sigma$ are given by \eqref{vars1}. In addition, $\delta\bnu$ and $\delta\bn$ are given as
\beq\label{nemvars2}
\delta\bnu=\partial_t\bxi+\bxi\times\bnu+\bu\cdot\nabla\bxi-\bkappa\cdot\nabla\bnu
,\qquad\qquad
\delta\bn=-\bkappa\cdot\nabla\bn+\bxi\times\bn.
\eeq
These variations reflect the Euler-Poincar\'e reduction from Lagrangian to Eulerian variables leading to \eqref{ELDisclvarprin}. In the absence of defects, $\bA$ is a pure gradient and the Euler-Poincar\'e reduction for liquid crystal dynamics has been presented in \cite{GBRa,GBRaTr2012,GBRaTr2013,Holm2002}. Here, we recall that, as before, the variational principle \eqref{ELDisclvarprin} is accompanied by the auxiliary equations
\beq\label{nemauxeqns}
\partial_t\sigma+\operatorname{div}(\sigma\bv)=0
,\qquad\qquad
\partial_t\rho+\operatorname{div}(\rho\bu)=0
,\qquad\qquad
\partial_t\bn+\bu\cdot\nabla\bn=\bnu\times\bn,
\eeq
which arise from the evolution laws 
\[
\sigma(\bx,t)={\int}\sigma_0(\bx_0)\delta(\bx-\boldsymbol\eta(\bx_0,t))\de^3x_0
\,,\qquad\quad
\rho(\bx,t)={\int}\rho_0(\bx_0)\delta(\bx-\boldsymbol\zeta(\bx_0,t))\de^3x_0
\]
and 
\[
\bn(\bx,t)=\chi(\bx_0,t)\bn_0(\bx_0)|_{\bx_0={\boldsymbol\zeta}^{-1}(\bx,t)}.
\]
Here, $\boldsymbol\eta(\bx_0,t)$ is the diffeomorphic Lagrangian path associated to the disclinations while  $\boldsymbol\zeta(\bx_0,t)$ is the corresponding path associated to the bulk fluid flow. Then, for example, the definition  \eqref{gammadef} is replaced by
\[
\gamma_{aj}=-\frac12\epsilon_{abc}\hat\gamma_{bcj}
,\qquad\text{where}\qquad
\hat\gamma_j=\big(( \chi(\partial_k\chi^{-1})+
\chi\hat{{\gamma}}_{k0}\chi^{-1})\circ{\boldsymbol\zeta}^{-1}\big)\partial_k\zeta_j,
\]
in compact notation.
The above evolution laws are accompanied by the defining relation
\[
\hat\nu=\dot\chi(\bx_0,t)\chi(\bx_0,t)^{-1}|_{\bx_0={\boldsymbol\zeta}^{-1}(\bx,t)},
\]
where 
the hat map can be inverted as in \eqref{nudef} to identify the associated angular velocity vector $\bnu$. These evolution laws for $\delta\hat\nu$ and $\bn$ recover exactly the relations \eqref{nemvars2}, where  $\hat\xi=(\delta\chi)\chi^{-1}\circ\boldsymbol\zeta^{-1}$.

Notice that the arbitrary variations $\delta\bmu$ yield the relation $
\bnu=J^{-1}{\bmu}/{\rho}+(\bA\cdot\bu-\Phi)\bn$, so that   transforming $\Phi\mapsto\Phi-\partial_t\varphi$ and $\bA\mapsto\bA+\nabla\varphi$, while leaving the momentum variable $\bmu$ unaffected, yields the gauge transformation rule for $\bnu$:
\beq\label{gaugenu2}
\bnu\mapsto\bnu +\bn (\partial_t+\bu\cdot\nabla)\varphi,
\eeq
thereby making  \eqref{ELDisclvarprin}  manifestly gauge invariant. In addition,  after recovering the first relation
in \eqref{charge&vel} from the arbitrary variations $\delta\Phi$, the variations $\delta\bA$ yield the expression of the disclination velocity
\beq\label{newvdef}
\bv=\bu+\frac1{\bmu\cdot\bn}\operatorname{curl}(\bv_\MH-\Delta^{-1}\bomega),
\eeq
which extends the previous flowless version \eqref{disclvel}. Then, after taking the variations $\delta\bv$ and $\delta\sigma$, the disclination density evolution is found by taking the curl of \eqref{disclpotevol} and  replacing \eqref{newvdef}, that is,
\beq\label{disclindensdynwflow}
\frac{\partial \bomega}{\partial t}={\operatorname{curl}}\bigg(\bu\times\bomega+\frac1{\mu_\|}\operatorname{curl}(\bv_\MH-\Delta^{-1}\bomega)\times\bomega\bigg),
\eeq
where we recall $\mu_\|=\bmu\cdot\bn$.
Also, taking the variations in \eqref{nemvars2} yields the analogue of \eqref{texteqsdef} in the presence of a hydrodynamic flow, that is,
\beq\label{texteqsdef-hyd}
\frac{\partial\bmu}{\partial t}+\operatorname{div}(\bu\bmu)=-\rho\bn\times\bh
-(\bomega\cdot\nabla) \bn
,\qquad\qquad
\frac{\partial\bn}{\partial t}+\bu\cdot\nabla\bn=\frac1{J\rho}\bmu\times\bn.
\eeq
Notice that the gauge charge \eqref{parmom} is now transported along the bulk flow:
\beq\label{gaugetransport}
\frac{\partial\mu_\|}{\partial t}+{\operatorname{div}}(\mu_\| \bu)=0.
\eeq

The fluid momentum equation is found by taking the variations $\delta\bu$ and $\delta\rho$ using \eqref{vars2}. After a few algebraic steps, this arises in the form
\[
m\rho\Big(\frac{\partial\bu}{\partial t}+\bu\cdot\nabla\bu\Big)=\mu_\|\bigg(\frac{\partial\bA}{\partial t} +\nabla\Phi-\bu\times\bomega\bigg)-\nabla p-\partial_i\bigg(\rho\nabla \bn\cdot\frac{\partial F}{\partial (\partial_i\bn)}\bigg)
+\bomega\times\operatorname{curl}\bv_\MH
\]
where we have used the first relation
in \eqref{charge&vel} together with \eqref{gaugetransport}. 
More importantly, we have made use of the following relation:
\begin{align}\nonumber
\partial_h \bn\cdot\frac{\delta }{\delta \bn}{\int}\bA\cdot\operatorname{curl}\bv_\MH\,\de^3x 
&\,=
\frac32\epsilon_{ijk}{\cal A}_i (\partial_h \bn)\cdot\partial_j\bn\times\partial_k \bn
-
    (\partial_h \bn)\cdot \bn\times(\bomega\cdot\nabla) \bn
\\
&=\,\nonumber
\frac32\epsilon_{ijk}{\cal A}_i  (\bn\times\bgamma_h)\cdot(\bn\times\bgamma_j)\times(\bn\times\bgamma_k)
-
    (\partial_h \bn)\cdot \bn\times(\bomega\cdot\nabla) \bn
\\
&=\,
\bomega\times\operatorname{curl}\bv_\MH,
\label{HelFunDer}
\end{align}
where the first equality follows from \eqref{varder} and the second equality from \eqref{gammaEvol}. Also, the third equality above is verified  using standard vector identities and expanding
\begin{align*}
2(\bomega\times\operatorname{curl}\bv_\MH)_h
=&\,
\epsilon_{abc}n_a\big(\bomega\times(\nabla n_a\times\nabla  n_c)\big)_{\!h}
\\=&\,
\epsilon_{abc}n_a\big((\bomega\cdot\nabla n_c)\partial_h n_b-(\bomega\cdot\nabla n_b)\partial_h n_c\big)
\\=&\,
-\bn\cdot(\bomega\cdot\nabla \bn)\times\partial_h \bn-\bn\cdot(\bomega\cdot\nabla \bn)\times\partial_h \bn
\\=&\,
-2\bn\cdot(\bomega\cdot\nabla \bn)\times\partial_h \bn
\\=&\,
-2(\partial_h \bn)\cdot \bn\times(\bomega\cdot\nabla \bn).
\end{align*}
Then, recalling the notation \eqref{vortcorr} 
and making use of \eqref{disclpotevol} and  \eqref{newvdef}, we have 
\beq\label{ELeqnwdisclin}
m\rho\Big(\frac{\partial\bu}{\partial t}+\bu\cdot\nabla\bu\Big)=-\nabla p+\bJ\times\bomega-\partial_i\bigg(\rho\nabla \bn\cdot\frac{\partial F}{\partial (\partial_i\bn)}\bigg),
\eeq
which closes the system comprised by \eqref{disclindensdynwflow}, \eqref{texteqsdef-hyd}, \eqref{ELeqnwdisclin}, and the mass transport equation
\beq\label{masstransp}
\frac{\partial\rho}{\partial t}+\operatorname{div}(\rho\bu)=0.
\eeq
We observe a direct analogy between equation \eqref{ELeqnwdisclin} and the Madelung momentum equation  \eqref{QFmom}, which however carries the quantum force $-\rho\nabla V_Q$ in place of the Frank energy terms $-\partial_i(\rho\nabla \bn\cdot{\partial F}/{\partial (\partial_i\bn)})$. The force term $\bJ\times\bomega$ identifies the  backflow effect generated by the action of the disclination field on the bulk flow. Also, the last  term in the disclination dynamics \eqref{disclindensdynwflow} retains the analogy with Hall magnetohydrodynamics already appeared for quantum fluids. The role of Hall transport in liquid crystals has recently been discussed in \cite{Coelho}.
We observe that, in the absence of disclinations, the trivial solution $\bomega=0$ of \eqref{disclindensdynwflow} allows to set $\mu_\|=0$, thereby returning Volovik's formulation of Ericksen-Leslie dynamics.

In turn, as already seen in the flowless case,  selecting the disclination line solution \eqref{disclinationline} leads to a formidable PDE system involving delta-like  forces and torques, so that suitable regularization techniques become necessary. 
We observe that the regularized Biot-Savart-nematodynamics model \eqref{vareqns2bar}-\eqref{MagnusDiscl} may  be extended to the hydrodynamic case by replacing $(\Phi,\bA)\rightsquigarrow(\bar\Phi,\barA)=(K*\Phi,K*\bA)$ everywhere in \eqref{ELDisclvarprin}, with the exception of the very last $\sigma$-term. Then, while the auxiliary equations \eqref{nemauxeqns} remain unaffected, taking variations of the resulting action principle leads to 
\beq\label{charge&velbar2}
\sigma=\overline{\mu_\|}
\,,\qquad\qquad
\bv=\frac1\sigma\overline{\big(\mu_\|\bu+\operatorname{curl}(\bv_\MH-\Delta^{-1}\bar\bomega)\big)}
,
\eeq
and
\begin{align*}
m\rho\Big(\frac{\partial \bu}{\partial t}+\bu\cdot\nabla\bu\Big)=&\,\mu_\|\Big(\frac{\partial \barA}{\partial t}+\nabla\bar\Phi\Big)
-(\mu_\|\bu+\operatorname{curl}\bv_\MH)\times\bar\bomega
-\nabla p-\partial_i\Big(\rho\nabla \bn\cdot\frac{\partial F}{\partial \bn_{,i}}\Big),
\end{align*}
along with 
\eqref{disclpotevol} and
${\partial_t\bmu}+\operatorname{div}(\bu\bmu)=\rho\bn\times\bh
+(\bar\bomega\cdot\nabla) \bn
$. As before, the first relation in \eqref{charge&velbar2} is preserved as the  Noether quantity associated to the invariance under the regularized gauge transformations \eqref{reggauge2}, with the third replaced by
$ 
\bnu\mapsto\bnu +\bn (\partial_t+\bu\cdot\nabla)\bar\varphi
$.
At this point, one introduces the disclination filament \eqref{disclinationline} in the curl of \eqref{disclpotevol}, that is $\partial_t\bomega=\operatorname{curl}(\bv\times\bomega)$, and  resorts to the Coulomb gauge. Then, since \eqref{disclpotevol} also leads to $\partial_t \barA+\nabla \bar\Phi=\overline{\bv\times \bomega}$, we are left with the  model equations
\begin{align*}
&\frac{\partial\rho}{\partial t}+\operatorname{div}(\rho\bu)=0,
\\
& m\rho\Big(\frac{\partial \bu}{\partial t}+\bu\cdot\nabla\bu\Big)= \mu_\|\overline{\bv\times \bomega}
-(\mu_\|\bu+\operatorname{curl}\bv_\MH)\times\bar\bomega
-\nabla p-\partial_i\Big(\rho\nabla \bn\cdot\frac{\partial F}{\partial \bn_{,i}}\Big),
\\
&\frac{\partial\bmu}{\partial t}+\operatorname{div}(\bu\bmu)=-\rho\bn\times\bh
-(\bar\bomega\cdot\nabla) \bn,
\qquad
\frac{\partial\bn}{\partial t}+\bu\cdot\nabla \bn 
=\frac\bmu{J\rho}\times\bn,
\\
&
\bR_s\times({\bR}_t-{\bv}(\bR))=0.
\end{align*}
Using the last equation, the regularized force exerted on the bulk can be written explicitly in terms of the disclination shape and velocity as follows:
\begin{multline*}
\mu_\|(\bx)\overline{\bv\times \bomega}(\bx)
-\big(\mu_\|(\bx)\bu(\bx)+\operatorname{curl}\bv_\MH(\bx)\big)\times\bar\bomega(\bx)
\\
=
n\pi{\int} K(\bx-\mathbf{R}(s))
\Big(
\mu_\|(\bx)\bigl(\mathbf{R}_t(s)-\mathbf u(\bx)\bigr)
-\operatorname{curl}\mathbf v_{\mathrm{MH}}(\bx)
\Big)\times \mathbf{R}_s(s)\de s.
\end{multline*}
Here, the first term is analogous to the expression obtained in \S\ref{sec:MVHD} in the context of quantum vortices and involves the relative motion of the filament and the fluid. Instead, the second term is the coupling between the filament and the nematic texture.
Again, the variational construction underlying this regularized model may lead to interesting prospects for the development of structure-preserving numerical schemes to study the interaction dynamics of singular disclination lines and the bulk flow in liquid crystal motion.


\subsection{Lagrangian and Hamiltonian formulations\label{sec:LiqxtalEqnsHam}}
The Lagrangian formulation of the system comprising the equations \eqref{disclindensdynwflow}-\eqref{texteqsdef-hyd} and \eqref{ELeqnwdisclin}-\eqref{masstransp} is obtained once again using the the Legendre transform relation, which in this case reads
\[
\bmu=J\rho\bnu+(\Phi-\bA\cdot\bu)\bn.
\]
Replacing this relation in \eqref{ELDisclvarprin} yields Hamilton's variational principle $\delta\int_{t_1}^{t_2}\!\ell\de t$, with the Lagrangian functional $\ell=\ell(\bu,\rho,\bnu,\bn,\bA,\,\,\dot{\!\!\boldsymbol{\cal A}},\Phi,\dot\Phi)$ given by
\begin{multline*}
\ell={\int}\bigg(\frac{m}2\rho|\bu|^2+
\frac{J}2\rho|\bnu+(\Phi-\bA\cdot\bu)\bn|^2+\sigma (\bA\cdot \bv-\Phi)
\\
-\rho U(\rho)
-\rho F(\bn,\nabla\bn)
-
\bA\cdot\operatorname{curl}\bv_\MH
-\frac{1}2(\operatorname{curl}\bA)\cdot(-\Delta^{-1})(\operatorname{curl}\bA)
\bigg)\de^3x,
\end{multline*}
and the variations given as in the previous section. Notice that the first line contains  kinematic and minimal-coupling terms, while the elastic and interaction energies are found in the second line along with the bulk fluid internal energy. Choosing the temporal gauge $\Phi=0$ and setting $\bA=0$  recovers the Euler-Poincar\'e Lagrangian $\ell=\ell(\bu,\rho,\bnu,\bn)$ which underlies Volovik's treatment for flowing liquid crystals  in the absence of disclinations \cite{Volovik}. As before, setting $\mu_\|=0$ in the latter case, recovers the well known Ericksen-Leslie hydrodynamics.

We now move on to presenting the Hamiltonian structure of the system given by \eqref{disclindensdynwflow}-\eqref{texteqsdef-hyd} and \eqref{ELeqnwdisclin}-\eqref{masstransp}.
Upon setting Weyl's gauge $\Phi=0$ and introducing the hydrodynamic momentum
\[
\bM=m\rho \bu+J\rho (\bn\cdot\bnu-\bA\cdot \bu)\bA
\]
along with the disclination momentum \eqref{vortmom},  the total energy Hamiltonian functional $\mathscr{H}=\mathscr{H}(\bM,\rho,\bmu,\bn,\bP,\sigma)$ is readily found to be
\begin{multline}\label{ELD-dinen}
\mathscr{H}={\int}\left(\frac1{2m\rho}\bigg|\bM+\frac{\bmu\cdot\bn}\sigma\bP\bigg|^2+\frac1{2J\rho}|\bmu|^2+\rho F(\bn,\nabla\bn)
+
\bomega\cdot\bv_\MH
+\frac{1}2\bomega(-\Delta^{-1})\bomega\right)\de^3 x.
\end{multline}
Then, the Hamiltonian structure is completed by the Poisson bracket functional $\{\mathscr{F},\mathscr{H}\}=\{\mathscr{F},\mathscr{H}\}(\bM,\rho,\bmu,\bn,\bP,\sigma)$ given by
\begin{multline}\label{ELDbkt}
\{\mathscr{F},\mathscr{H}\}=
{\int}\bM\cdot\bigg(\frac{\delta \mathscr{F}}{\delta \bM}\cdot\nabla\frac{\delta \mathscr{H}}{\delta \bM}-\frac{\delta \mathscr{H}}{\delta \bM}\cdot\nabla\frac{\delta \mathscr{F}}{\delta \bM}\bigg)\de^3 x
-
{\int}\rho\bigg(\frac{\delta \mathscr{F}}{\delta \bM}\cdot\nabla\frac{\delta \mathscr{H}}{\delta \rho}-\frac{\delta \mathscr{H}}{\delta \bM}\cdot\nabla\frac{\delta \mathscr{F}}{\delta \rho}\bigg)\de^3 x
\\
+{\int}\bigg[\bmu\cdot\frac{\delta \mathscr{F}}{\delta \bmu}\times\frac{\delta \mathscr{H}}{\delta \bmu}+\bn\cdot\bigg(\frac{\delta \mathscr{F}}{\delta \bn}\times\frac{\delta \mathscr{H}}{\delta \bmu}-\frac{\delta \mathscr{H}}{\delta \bn}\times\frac{\delta \mathscr{F}}{\delta \bmu}\bigg)\bigg]\de^3 x
\\
-
{\int}\bmu\cdot\bigg(\frac{\delta \mathscr{F}}{\delta \bM}\cdot\nabla\frac{\delta \mathscr{H}}{\delta \bmu}-\frac{\delta \mathscr{H}}{\delta \bM}\cdot\nabla\frac{\delta \mathscr{F}}{\delta \bmu}\bigg)\de^3 x
-
{\int}\bn\cdot\operatorname{div}\bigg(\frac{\delta \mathscr{F}}{\delta \bM}\frac{\delta \mathscr{H}}{\delta \bn}-\frac{\delta \mathscr{H}}{\delta \bM}\frac{\delta \mathscr{F}}{\delta \bn}\bigg)\de^3 x
\\
+
{\int}\bP\cdot\bigg(\frac{\delta \mathscr{F}}{\delta \bP}\cdot\nabla\frac{\delta \mathscr{H}}{\delta \bP}-\frac{\delta \mathscr{H}}{\delta \bP}\cdot\nabla\frac{\delta \mathscr{F}}{\delta \bP}\bigg)\de^3 x
-
{\int}\sigma\bigg(\frac{\delta \mathscr{F}}{\delta \bP}\cdot\nabla\frac{\delta \mathscr{H}}{\delta \sigma}-\frac{\delta \mathscr{H}}{\delta \bP}\cdot\nabla\frac{\delta \mathscr{F}}{\delta \sigma}\bigg)\de^3 x,
\end{multline}
where we have used the notation $\operatorname{div}(\bv\bw)=\partial_j(v_j \bw)$. Here, the first three lines comprise the Poisson bracket structure of liquid crystal hydrodynamics in Volovik's formulation; see \cite{GBRa} for further details on its geometric structure involving nested semidirect-products of Lie groups. The fourth line is the defect bracket which already appeared in  \S\ref{sec:MCHDHamSt} within the context of quantum vortices.

Other than the  vortex helicity invariant in \eqref{disclhel}, the most interesting feature of the Poisson bracket \eqref{ELDbkt} is the appearance of hydrodynamic helicity as an additional Casimir invariant. This reads
\[
{\int}\boldsymbol{C}\cdot\operatorname{curl}\boldsymbol{C}\,\de^3x
\,,\qquad\text{with}\qquad
\boldsymbol{C}=m\bu+\frac1{\rho}\big(\bmu\cdot\bn\times\nabla\bn-{\mu_\|}\bA\big).
\]
This helicity invariant extends a previous result by the authors \cite{GBTr2010}, obtained in the absence of disclinations. Indeed, in the latter case one sets $\bA=0$ and the reversible Hamiltonian evolution of nematic liquid crystals preserves the above helicity functional. As shown in \cite{GBTr2010}, this type of hydrodynamic helicity can be found for various cases of condensed matter dynamics with symmetry breaking.

\section{Conclusions and perspectives}

Blending the Euler-Poincar\'e variational approach to continuum dynamics with the gauge theory of topological defects, we have provided a unified framework for the interaction dynamics of topological defects and the evolving background fields. Under the assumption of ideal conservative evolution in the absence of topological changes, we have used the standard minimal coupling approach in gauge theory and applied our modeling platform to vortex motion in quantum fluids and disclination dynamics in uniaxial nematic liquid crystals. The defect-defect interaction energy is realized in the different contexts via the Biot-Savart vortex energy from Euler's incompressible flows, as proposed by Kleinert \cite{Kleinert,Kleinert1} in the context of superfluid dynamics. While the proposed approach is based on older work of Dirac and Takabayasi \cite{Takabayasi}, a more modern analogy also appeared  with the Hall magnetohydrodynamics theory of magnetized plasmas, where the role of the defect density $\bomega$ is played by the magnetic field $\bB$ while the defect current $\bJ=\operatorname{curl}(-\Delta^{-1})\bomega$ is the analogue of the plasma current $\mathbf{J}=\operatorname{curl}\bB$. This analogy is reflected in the variational and Hamiltoniastructure of the different models considered here.

In the case of Gross-Pitaevskii vortex dynamics, the construction extends the Dirac-Ta\-ka\-ba\-ya\-shi
 theory to the case of a time evolving vortex density by resorting to simple minimal coupling within the Dirac-Frenkel action principle for quantum wavefunction evolution. When the vortex density is singular, the resulting Biot-Savart-Schr\"odinger system requires a regularization due to well-known divergences also appearing in incompressible fluid dynamics.  Given its underlying variational structure, the regularized Biot-Savart-Schr\"odinger model opens new perspectives for structure preserving numerical schemes for the simulation of the vortex-wavefunction interaction. In  Madelung's  hydrodynamic setting, one obtains a quantum analogue of Hall magnetohydrodynamics in which, similarly to Biot-Savart-Schr\"odinger dynamics,  the vortex density has a nontrivial relative motion with respect to the bulk flow. This relative motion is associated to the Biot-Savart vortex energy and produces a reaction force $\bJ\times\bomega$ on the quantum fluid. 

After treating quantum vortex dynamics, the discussion shifted towards disclinations in nematic media.
Before tackling nematic liquid crystals, we approached  disclination dynamics in flowless nematic textures. In dealing with this case, we addressed an apparent ambiguity  in the works of Kl\'eman \cite{Kleman}, Volovik \cite{Volovik}, and Halperin \cite{Halperin} about the definition of disclination density. In particular, we have established the unique relation among the different definitions by showing that the curvature of the non-Abelian gauge potential \eqref{wrynessdef} for the orientational nematic order is necessarily parallel to the director field. Importantly, this parallel component  identifies a different curvature, this time associated to Abelian planar rotations around the director, and whose vector potential $\bA$ in \eqref{connform} involves the Mermin-Ho velocity $\bv_\MH$ in \eqref{Mermin_Ho}. While related, these latter two quantities remain distinctively different. The Mermin-Ho velocity plays a crucial in realizing the disclination-texture interaction energy as $\int \bomega\cdot\bv_\MH\de^3 x$. 

After clarifying the definition of the disclination density and its vector potential, we have applied again minimal coupling in such a way to couple disclination density dynamics to the evolution of the background nematic texture. In doing this, we have resorted to Volovik's extension of Ericksen-Leslie theory, the latter being otherwise  incompatible with disclination dynamics. We have clarified that this incompatibility is due to the fact that the parallel angular momentum $\mu_\|=\bmu\cdot\bn$ is set to zero in Ericksen-Leslie theory, while it remains unspecified upfront in Volovik's extension. In general, we have shown that disclination dynamics can be captured in Volovik's model whenever $\mu_\|$ is strictly sign-definite. In the absence of a fluid flow, the disclination velocity carries two contributions: one is the Peach-Koehler term arising from the intrinsic (Biot-Savart) disclination energy while the other involves the Mermin-Ho velocity and is associated to the direct interaction between the disclinations and the texture. This relative motion feeds back in the nematic angular momentum dynamics via a torque term that is proportional to the disclination density $\bomega$ itself.  When the latter is given by a singular disclination line, a suitable regularization procedure leads to a  Biot-Savart nematodynamics model for the coupled dynamics. In this case, the Rasetti-Regge bracket of quantum vortex dynamics applies directly to the motion of disclination lines, thereby establishing an interesting analogy whose geometric properties will be explored elsewhere.

At last, we considered the disclination dynamics in flowing liquid crystals. In this case, the approach to textures can be extended by the addition of the minimal coupling between the disclinations and the bulk fluid flow while leaving the rest of the construction unchanged. In this case, the relative disclination velocity with respect to the bulk is given by the velocity expression previously found in the absence of flow. The Peach-Koehler term associated to the Biot-Savart disclination energy generates a reaction on the bulk whose expression $\bJ\times\bomega$ is identical to the corresponding term already found in the case of quantum vortices. Then, if the disclination density is given by a single defect filament this reaction force becomes singular thereby requiring a suitable regularization which in our case arises again from a variational principle. Finally, the Hamiltonian structure of the overall continuum model reveals the coexistence of two dynamical invariants: the disclination helicity and the hydrodynamic helicity, thereby extending previous results obtained in the absence of defects \cite{GBTr2010}.

In the future, we would like to understand how to go beyond the ideal limit by adding dissipative irreversible effects. The latter are most important in the case of liquid crystals, for which realistic situations involve damped or overdamped dynamics. For example, the addition of dissipative terms can be realized in the present variational setting by resorting to Rayleigh dissipation \cite{GBRaTr2012,Virga}. Moreover, we would also like to understand how to account for the reconnection of defect lines. In this respect, the analogies with plasma theory can be beneficial as already noticed in \cite{Narita}. In general, reconnection is allowed by breaking the frozen-in condition $\partial_t\bomega-\operatorname{curl}(\bv\times\bomega)=0$, which would then involve a  source term on the right-hand side. This source term can be either conservative, as in e.g., collisionless magnetic reconnection \cite{Narita}, or diffusive, as in e.g., resistive magnetohydrodynamics.  The first option may involve accounting for a time-evolving stress tensor into the vortex dynamics. In the case of quantum vortices, this would mimic the effects of the so called \emph{quantum stress tensor} $\Bbb{T}_{jk}=-(\hbar/m)^2\rho\partial_j\partial_k\ln\rho/4$, while in the case of nematic media this would correspond to the effects the \emph{Ericksen stress tensor} $\Bbb{T}_{jk}=-\rho\bn_{,k}\cdot \partial F/\partial \bn_{,j}$. Alternatively, dissipative  contributions to defect reconnection can be directly realized by invoking a nonzero defect diffusivity $D$ to write $\partial_t\bomega=\operatorname{curl}(\bv\times\bomega)+D\Delta\bomega$.

A separate comment is required about possible analogies with the $Q$-tensor theory of liquid crystals \cite{deGennes}, which allows the simultaneous treatment of uniaxial and biaxial phases and is currently widely used to study defect dynamics \cite{Angheluta,Ronning,Schimming}. Different formulations of $Q$-tensor dynamics are available \cite{Close,QiSh1998,VK1981}, with the most popular being the Beris-Edwards variant \cite{BeEd1994}. While the perspective of extending the present construction to $Q$-tensor evolution would be appealing, the absence of a well-defined Mermin-Ho velocity in the corresponding setting seems to prevent this possibility. Indeed, the extension of the present construction to the biaxial phase seems to require advancing two separate director fields altogether. 

A further future direction  involves extending the present construction to the dynamics of defects within complex fluids with multipolar interactions  \cite{Tr2012}, for which the order parameter field is replaced by a probability density on the finite-dimensional order parameter space. This and the other directions above will be explored in future work.

\paragraph{Acknowledgments.} We are greatly indebted to Michael Foskett, Darryl Holm, and Mauro Spera for several enlightening discussions on this and related topics over the years. FGB and CT acknowledge financial support by the Leverhulme Trust Research Project Grant RPG-2023-078. Also, TSR and CT are grateful to the Division of Mathematical and Physical Sciences at NTU Singapore for their kind hospitality during the course of this work.

\begingroup
\linespread{0.888}\selectfont 

\endgroup


\begin{thebibliography}{99}\small

\bibitem{Aharonov}
Aharonov, Y; Bohm, D. {\it Significance of electromagnetic potentials in quantum theory}. Phys. Rev. 115  (1959), 485-491.

\bibitem{Angheluta}
Angheluta, L.; Chen, Z.; Marchetti, M.C.; Bowick, M.J. {\it  The role of fluid flow in the dynamics of active nematic defects}. New J. Phys.  23 (2021), 033009.

\bibitem{BeEd1994}
Beris A. N. and B. J. Edwards, \textit{Thermodynamics of Flowing Systems}
Oxford University Press, Oxford, 1994.

\bibitem{Berry}
Berry, M.V. {\it Quantal phase factors accompanying adiabatic changes}. {Proc. R. Soc. A} 392, (1984), 45-57.

\bibitem{Berry1}
Berry, M.V.; Dennis, M.R. {\it Reconnections of wave vortex lines}.  Eur. J. Phys. 33 (2012), 723-731.

\bibitem{Bohm}
Bohm, D.A. {\it Suggested interpretation of the quantum theory in terms of ``hidden" variables, I}. {Phys. Rev.} (85), (1952) 166-179. 


\bibitem{Birula}
Bialynicki-Birula, I.; Bialynicka-Birula, Z.; \'Sliwa, C. {\it Motion of vortex lines in quantum mechanics}. {Phys. Rev. A} 61 (2000), 32-110.


\bibitem{Nazarenko}
Bustamante, M.D.; Nazarenko, S. {\it Derivation of the Biot-Savart equation from the nonlinear Schr\"odinger equation}. Phys. Rev. E 92 (2015), 053019.

\bibitem{Carpio}
 Carpio, A.; Chapman, S.J.; Howison, S.D.; Ockendon, J.R. {\it Dynamics of line singularities}. Phil. Trans. R. Soc. A  355 (1997), 2013-2024.

\bibitem{Cermelli}
 Cermelli, P.; Fried, E. {\it  The evolution equation for a disclination in a nematic liquid crystal}. Proc. R. Soc. A  458 (2002), 1-20.
 
 \bibitem{Close}
 Close, A.R.D.; Tronci, C. {\it  Equivalent variational approaches to biaxial liquid crystal dynamics}. Proc. R. Soc. A 471 (2015), 20150308.
 
\bibitem{Coelho}
Coelho, R.C.V.; Zhao, H.; Amaral, G.N.C.; Smalyukh, I.I.; Telo da Gama, M.M.; Tasinkevych, M. {\it ``Hall'' transport of liquid crystal solitons in Couette flow}. {\tt arXiv:2405.10850}.
 
 \bibitem{deGennes}
de Gennes, P. G.; Prost, J. The Physics of Liquid Crystals. Oxford University Press. 1993.
 
\bibitem{Dirac}
Dirac, P.A.M. {\it Quantised singularities in the electromagnetic field}. Proc. R. Soc. A 133 (1931), 60-72.

\bibitem{Volovik}
Dzyaloshinskii, I.E.; Volovick, G.E. {\it Poisson brackets in condensed matter physics}. Ann. Phys. 125 (1980), 67-97.

\bibitem{Eringen}
Eringen, A.C. {\it A unified continuum theory of electrodynamics of liquid crystals}. Int. J. Engng. Sci.  35 (1997), 1137-1157.


\bibitem{Fischer}
 Fischer, U.R. {\it Motion of quantized vortices as elementary objects}. Ann. Phys.  278 (1999),  62-85.
 
 \bibitem{Ricca}
Foresti, M.; Ricca, R.L. {\it Hydrodynamics of a quantum vortex in the presence of twist}. J. Fluid Mech. 904 (2020), A25 1-11.

\bibitem{FoHoTr19}
Foskett, M.S.; Holm, D.D.; Tronci, C.  {\it Geometry of nonadiabatic quantum hydrodynamics}. {Acta Appl. Math.} 162 (2019), 63-103.


\bibitem{FoTr}
Foskett, M.S.; Tronci, C. {\it  Holonomy and vortex structures in quantum hydrodynamics}. In ``Hamiltonian Systems: Dynamics, Analysis, Applications''.  Fathi, A.; P. J. Morrison,P.J.; Seara, T.M.; and Tabachnikov, S. (eds.).  Math. Sci. Res. Inst. Pub. 72, 173-214. Cambridge University Press.   2024. 

\bibitem{GBRa}
Gay-Balmaz, F.; Ratiu, T.S. {\it The geometric structure of complex fluids}. {Adv. Applied Math.}
42, (2009), 176-275.


\bibitem{GBRaTr2013}
Gay-Balmaz, F.; Ratiu, T.S.; Tronci, C. {\it Equivalent theories of liquid crystal dynamics}.
{Arch. Rational Mech. Anal.} 210 (2013), 773-811.

\bibitem{GBRaTr2012}
Gay-Balmaz, F.; Ratiu, T.S.; Tronci, C. {\it Euler-Poincar\'e approaches to nematodynamics}.
{Acta Appl. Math.} 120 (2012), 127-151.


\bibitem{GBTr2010}
Gay-Balmaz, F.; Tronci, C. {\it Reduction theory for symmetry breaking with applications to nematic systems}. {Phys. D} 239 (2010), 1929-1947.

\bibitem{Halperin}
Halperin, B.I. {\it Statistical mechanics of topological defects}. In Physics of Defects, eds. R.
Balian, et al, North-Holland, N.Y. 1981.



\bibitem{Holm2002}
Holm, D.D. {\it Euler-Poincar\'e dynamics of perfect complex fluids.} In ``Geometry, Mechanics, and Dynamics''. Newton, P.; Holmes, P.; Weinstein, A. (eds.)  169-180. Springer, 2002.

\bibitem{Holm}
Holm, D.D. {\it Hall magnetohydrodynamics: conservation laws and Lyapunov stability}. Phys. Fluids 30 (1987), 1310-1322.


\bibitem{Holm03}
Holm, D.D. {\it Rasetti-Regge Dirac bracket formulation of Lagrangian fluid dynamics of vortex filaments}. {Math. Comp. Simulation} 62  (2003), 53-63.


\bibitem{HoMaRa98}
Holm, D.D., Marsden, J.E.; Ratiu, T.S. {\it The Euler-Poincar\'e equations and semidirect products with applications to continuum theories}. { Adv. Math.} 137  (1998), 1-81.


\bibitem{Barenghi}
Idowu, O.C.; Kivotides,   D.; Barenghi, C.F.; Samuels,  D. C  {\it Equation for self-consistent superfluid vortex line dynamics}. J.  Low Temp. Phys. 120  (2000), 269-280.

\bibitem{Ilgisonis}
 Ilgisonis, V.I.; Lakhin,  V.P. {\it Lagrangean structure of hydrodynamic plasma models and conservation laws}. Plasma Phys. Rep. 25 (1999), 58-69.


\bibitem{Kleinert}
Kleinert, H. {Multivalued Fields in Condensed Matter, Electromagnetism, and Gravitation.} World Scientific. 2008.

\bibitem{Kleinert1}
Kleinert, H. {\it Theory of fluctuating nonholonomic fields and applications: Statistical mechanics of vortices and defects and new physical laws in spaces with curvature and torsion}.  In ``Formation and Interactions of Topological Defects''. Davis, A.C. and Brandenberger, R. (eds.). NATO ASI Series, 349 (1995), 201-232. Springer, Boston, MA.

\bibitem{Kleman}
Kl\'eman, M. {\it Defect densities in directional media, mainly liquid crystals}. Phil. Mag. 27 (1973), 1057-1072.


\bibitem{Koch}
Koch, C-M.; Wilczek, M. {\it Role of advective inertia in active nematic turbulence}. Phys. Rev. Lett.  127 (2021), 268005.


\bibitem{Mead}
Mead, C. A. {\it The geometric phase in molecular systems.} {Rev. Mod. Phys.} 64 (1992),  51.

\bibitem{MerminHo}
Mermin, N.D.; Ho, T.L. {\it Circulation and angular momentum in the a phase of superfluid Helium-3}. {Phys. Rev. Lett.} 36 (1976), 594-597.


\bibitem{Madelung}
Madelung, E. {\it Quantentheorie in hydrodynamischer Form}. {Z. Phys.} 40, (1927), 
322-326.

\bibitem{MaWe83}
 Marsden, J.; Weinstein, A. {\it  Coadjoint orbits, vortices, and Clebsch variables for incompressible fluids}. Phys. D 7 (1983),  305-323.

 
\bibitem{Narita}
Narita, Y.; Baumjohann, W. {\it  Lessons on collisionless reconnection from quantum fluids}. Front. Phys. 2 (2014), 76 1-11.


\bibitem{Ohsawa}
Ohsawa, T. {\it Relative dynamics of vortices in confined Bose–Einstein condensates}. Phys. D 476 (2025), 134639, 1-10


\bibitem{Peach}
Peach, M.; Koehler, J.S.; {\it The forces exerted on dislocations and the stress fields produced by them}. Phys Rev.  80 (1950), 436-439.

\bibitem{PennaSpera1989}
Penna, V.; Spera, M. {\it A geometric approach to quantum vortices}. {J. Math. Phys.} 30 (1989),  2778-2784.



\bibitem{QiSh1998}
Qian T.  and P. Sheng, {\it Generalized hydrodynamic equations for nematic liquid crystals}. Phys. Rev. E 58 (1998), 7475-7485.

\bibitem{RasettiRegge}
Rasetti, M. and Regge, T. {\it Vortices in He II, current algebras and quantum knots}. {Phys. A} 80 (1975), 217-233.

\bibitem{RaTr20}
Rawlinson, J.I.; Tronci, C. {\it Regularized Born-Oppenheimer molecular dynamics}. Phys. Rev. A 102 (2020), 032811.

\bibitem{Berloff}
Roberts, P.H.; Berloff, N.G. {\it The nonlinear Schrödinger equation as a model of superfluidity}. In ``Quantized Vortex Dynamics and Superfluid Turbulence''. Barenghi, C.F.; Donnelly, R.J.; Vinen, W.F. (eds.).  Lect. Notes  Phys.  571.  235–257. Springer. 2001.


\bibitem{Ronning}
 R\o nning, J.; Marchetti, C.M; Bowick, M.J.; Angheluta, L. {\it Flow around topological defects in active nematic films}.
Proc. R. Soc. A 478 (2022), 20210879.

\bibitem{Ruban}
Ruban, V.P. {\it Dynamics of straight vortex filaments in a Bose-Einstein condensate with the Gaussian density profile}. JETP 124 (2017), 932-942.

\bibitem{Schimming}
Schimming, C.D.;  Vi\~nals, J. {\it Kinematics and dynamics of disclination lines in three-dimensional nematics}. Proc. R. Soc. A  479 (2023), 20230042


\bibitem{Schimming2}
Schimming, C.D.;  Vi\~nals, J.  {\it  Singularity identification for the characterization of topology, geometry, and motion of nematic disclination lines}. Soft Matter  18 (2022), 2234-2244.

\bibitem{Sonin}
Sonin, E. {Magnus force in superfluids and superconductors}. Phys. Rev. B 55 (1997), 485-501.


\bibitem{Virga}
 Sonnet, A.M.; Virga, E.G. {\it Dynamics of dissipative ordered fluids}. Phys. Rev. E 64 (2001), 031705.

\bibitem{Takabayasi}
Takabayasi, T. {\it Hydrodynamical formalism of quantum mechanics and Aharonov-Bohm effect}. Prog. Theor. Phys. 69 (1983), 1323-1344.


\bibitem{Tr2012}
Tronci, C. {\it  Hybrid models for complex fluids with multipolar interactions}. J. Geom. Mech. 4 (2012), 333-363

\bibitem{Tr2026}
Tronci, C. {\it  Madelung hydrodynamics of spin-orbit coupling: action principles, currents, and correlations}. Proc. R. Soc. A (submitted). {\tt arXiv:2601.10698}

\bibitem{VK1981}
Volovik, G.E.; Kats,   E.I. {\it Nonlinear hydrodynamics of liquid crystals}.   Sov. Phys.,  54 (1981), 122-126.


\bibitem{Wyatt}
Wyatt, R.E. {Quantum Dynamics with Trajectories: Introduction to Quantum Hydrodynamics.} Springer. 2005.

\bibitem{Zarnescu}
Zarnescu, A. {\it Mathematical problems of nematic liquid crystals: between dynamical and stationary problems}. Phil. Trans. R. Soc. A  379 (2021), 20200432.

\bibitem{Zuccher}
 Zuccher, S.; Caliari, M.; Baggaley, A.W.; Barenghi, C.F. {\it  Quantum vortex reconnections}. Phys. Fluids 24  (2012),  125108.

\end{thebibliography}
\end{document}